\documentclass[10pt,amsmath,amssymb,aps,pra,twocolumn,longbibliography,floatfix,superscriptaddress,nofootinbib]{revtex4-1}

\usepackage[utf8]{inputenc}
\usepackage[T1]{fontenc}

\usepackage{dsfont}
\usepackage{times}
\usepackage{txfonts}

\usepackage{physics}
\usepackage{mathtools}
\usepackage{mleftright}\mleftright
\usepackage{float}
\usepackage{enumitem}
\usepackage{bm}

\usepackage{algorithm}
\usepackage{algpseudocode}
\floatstyle{boxed}
\newfloat{algorithm}{t}{lop}
\floatname{algorithm}{Algorithm} 

\usepackage{multirow}
\usepackage{diagbox}
\usepackage{soul}
\floatstyle{boxed}
\usepackage{dcolumn}

\usepackage{graphicx}
\usepackage{xcolor}

\usepackage{tikz}
\usetikzlibrary{decorations.pathmorphing, fit}

\usepackage[
final,
colorlinks,
allcolors = blue
]{hyperref}

\newcommand{\rp}[1]{(\ref{#1})}
\newcommand{\pt}[1]{\left( #1 \right)}
\newcommand{\pq}[1]{\left[ #1 \right]}
\newcommand{\pg}[1]{\left\{ #1 \right\}}
\newcommand{\ii}{{\rm i}}
\newcommand{\nn}{{\nonumber}}

\newcommand{\da}{^\dagger}
\newcommand{\oo}[0]{^{\circ}}

\definecolor{blue}{rgb}{0,0,0.8}

\definecolor{red}{rgb}{0.8,0,0}

\definecolor{green}{rgb}{0,0.6,0}

\usepackage[normalem]{ulem}
\newcommand{\stkout}[1]{\ifmmode\textrm{\sout{\ensuremath{#1}}}\else\sout{#1}\fi}

\begin{document}
\title{Can gravity mediate the transmission of quantum information?}
\author{Andrea Mari}
\affiliation{Physics Division, School of Science and Technology, Universit\`a di Camerino, 62032 Camerino, Italy}
\author{Stefano Zippilli}
\affiliation{Physics Division, School of Science and Technology, Universit\`a di Camerino, 62032 Camerino, Italy}
\author{David Vitali}
\affiliation{Physics Division, School of Science and Technology, Universit\`a di Camerino, 62032 Camerino, Italy}
\affiliation{INFN, Sezione di Perugia, I-06123 Perugia, Italy}
\affiliation{CNR-INO, L.go Enrico Fermi 6, I-50125 Firenze, Italy}

\begin{abstract}
We propose an experiment to test the non-classicality of the gravitational interaction. We consider two optomechanical systems that are perfectly isolated, except for a weak gravitational coupling. If a suitable resonance condition is satisfied, an optical signal can be transmitted from one system to the other
over a narrow frequency band, a phenomenon that we call  {\it gravitationally induced transparency}. In this framework, the challenging problem of testing the quantum nature of gravity is mapped to the easier task of determining the non-classicality of the gravitationally-induced optical channel: If the optical channel is not entanglement-breaking, then gravity must have a quantum nature.
This approach is applicable without making any assumption on the, currently unknown, correct model of gravity in the quantum regime. In the second part of this work, we model gravity as a quadratic Hamiltonian interaction (e.g.~a weak Newtonian force),  resulting in a Gaussian thermal attenuator channel between the two systems. Depending on the strength of thermal noise, the system presents a sharp transition from an entanglement-breaking
to a non-classical channel capable not only of entanglement preservation but also of asymptotically perfect quantum communication.
\end{abstract}

\maketitle


One of the main open problems in physics is merging quantum mechanics and gravity into a unified and consistent theory \cite{woodard2009far, kiefer2007quantum}.
Today, despite many interesting attempts in this direction \cite{ashtekar2021short, schwarz1999string}, the problem is unsolved and it is even debated whether gravity requires a quantum description in the first place \cite{carlip2008quantum, tilloy2019does, oppenheim2023postquantum}. In fact, theoretical models have been proposed in which gravity behaves as a fundamentally classical entity, even in the presence of quantum source masses \cite{kafri2013noise, kafri2014classical, kafri2015bounds, tilloy2016sourcing, oppenheim2023postquantum, oppenheim2023gravitationally,tilloy2024general, hu2008stochastic}.

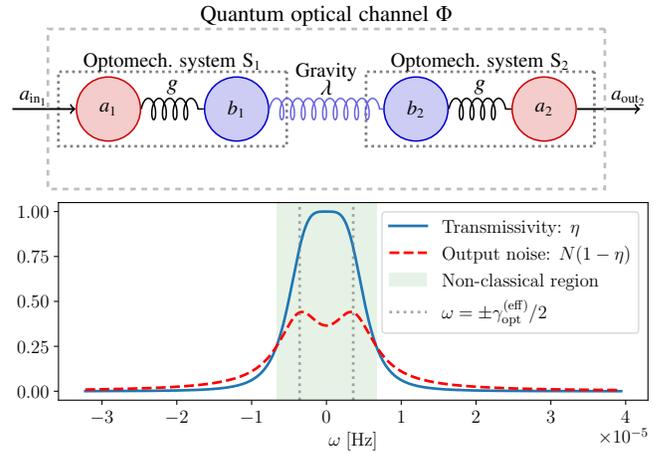
\begin{figure}[h!]
\resizebox{\columnwidth}{!}{
    \begin{tikzpicture}[thick,scale=1.2, every node/.style={scale=1.2}]
        \node[circle, draw=red, fill=red!20, thick, minimum size=1cm] (a1) at (0,0) {$a_1$};
        \node[circle, draw=blue, fill=blue!20, thick, minimum size=1cm] (b1) at (2,0) {$b_1$};
        \node[draw, dotted, gray, thick, line width=0.5mm, fit=(a1)(b1), inner sep=-1pt] {}; 
        \node[circle, draw=red, fill=red!20, thick, minimum size=1cm] (a2) at (6.8,0) {$a_2$};
        \node[circle, draw=blue, fill=blue!20, thick, minimum size=1cm] (b2) at (4.8,0) {$b_2$};
         \node[draw, dotted, gray, thick, line width=0.5mm, fit=(a2)(b2), inner sep=-1pt] {}; 
        \draw[->, line width=1pt] (-1.5,0) -- (a1) node[midway, above=-2pt] {$a_{\mathrm{in}_1} \quad$};
        \draw[->, line width=1pt] (a2) -- (8.3,0)  node[midway, above=-2pt] {$\quad\quad a_{\mathrm{out}_2}$};
        \draw[decorate,decoration={coil,aspect=0.4,segment length=2mm,amplitude=2mm}] (a1) -- (b1)
            node[midway, above=3pt, text width=2.8cm, align=center] {\small Optomech.~system~S$_{\rm 1}$  \\ \vspace{-0mm} \normalsize $g$};
        \draw[decorate,decoration={coil,aspect=0.4,segment length=2mm,amplitude=2mm}] (b2) -- (a2)
            node[midway, above=3pt, text width=2.8cm, align=center] {\small Optomech.~system~S$_{\rm 2}$  \\ \vspace{-0mm} \normalsize  $g$};
        \draw[decorate,decoration={coil, aspect=0.4,segment length=2mm, amplitude=2mm},  draw=blue!60] (b1) -- (b2)
            node[midway, above=3pt, text width=7.8cm, align=center, 
            ] {\normalsize Quantum optical channel $\Phi$\\
            \vspace{0.5 cm} \small Gravity \\ \vspace{-1mm}  \normalsize   $\lambda$};    
        \node[draw, lightgray, dashed, thick, line width=0.5mm, fit=(a1)(a2), inner sep=-10pt, minimum height=2.5 cm]{};  
    \end{tikzpicture}
}
\includegraphics[width=1.0\linewidth]{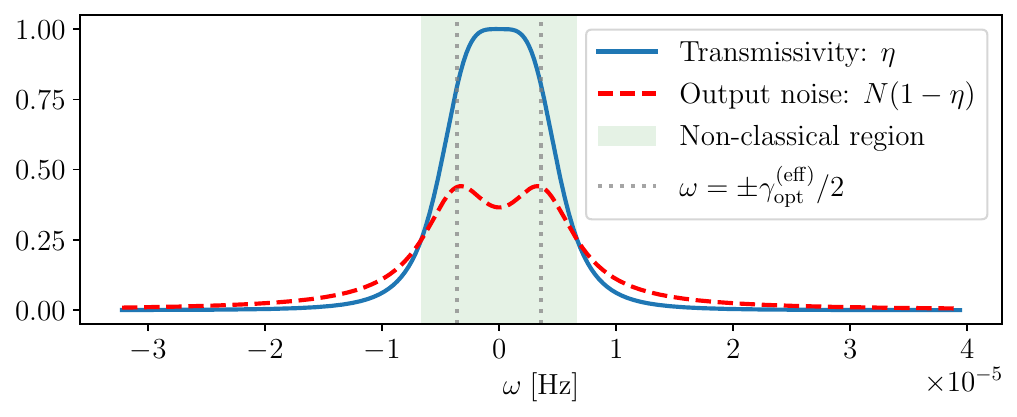}
\caption{(Top) The apparatus required to observe the {\it gravitationally-induced transparency} (GIT) phenomenon is based on two opto-mechanical systems, $S_1$ and $S_2$. Each system is composed of an optical cavity mode $a_j$ that is coupled with a mechanical resonator $b_j$, where $j=1,2$. By tuning the mode frequencies and the optomechanical coupling $g$, gravity can induce an effective quantum channel $\Phi$ from the optical input of $S_1$ to the optical output of $S_2$. Any quantum optics experiment demonstrating that $\Phi$ is not {\it entanglement-breaking} would imply that gravity is a non-classical phenomenon.
(Bottom) Effective transmissivity and output noise as a function of the input probe frequency $\omega$ (relative to the cavity frequency). 
When the transmissivity is higher than the output noise, the channel $\Phi$ is non-classical.
This plot is based on the quadratic gravitational interaction defined in Eq.~\eqref{eq:interaction} and on the following physical parameters: 
$\omega_B=2 \pi\times0.03$ Hz, $\gamma=10^{-14} \omega_B$, $\kappa=0.1 \omega_B$, $\lambda=3.58\times 10^{-6}$ Hz, $g=1.84 \times 10^{-4}$ Hz, and $T=1$ mK.
For more details, see Fig. \ref{fig:git} in Supplemental Material  \cite{supplemental_material} \ref{app:full_solution}.}
\label{fig:scheme}
\end{figure}

In this work, we use a quantum communication theory \cite{khatri2020principles} approach and propose an experimental protocol to test whether gravity is a classical or a quantum phenomenon. Specifically, the experiment is designed to falsify the hypothesis that gravity can be described as a classical process locally acting on the quantum dynamics of test masses.
The apparatus, schematically represented in Fig.~\ref{fig:scheme}, is based on two opto-mechanical systems \cite{milburn2011introduction, aspelmeyer2014cavity, barzanjeh2022optomechanics} laser-driven in their steady-state regime and isolated from each other, apart from a weak gravitational interaction acting as a mediator between their mechanical resonators. A suitable resonance condition can be tuned such that gravity induces an effective optical communication channel between the two systems. We call this phenomenon {\it gravitationally-induced transparency} (GIT). In this setting, the general question of whether gravity is classical or quantum is mapped to the well-defined problem of determining the non-classical nature of an optical channel.

Once the problem is reduced to the characterization of a quantum optical channel, we can use standard tools from quantum information theory to distinguish intrinsically quantum processes from classical processes. A well-established criterion is the notion of {\it entanglement-breaking} channels \cite{horodecki2003entanglement, holevo2008entanglement}.
A channel $\Phi$ is entanglement-breaking if it always produces a separable state $\rho_{\rm out}^{(\rm AB)}
=(\mathbb{I} \otimes \Phi)(\rho_{\rm in}^{(\rm AB)})$ when applied to a sub-system of a composite  system initially prepared in
any entangled state $\rho_{\rm in}^{(\rm AB)}$.
Crucially, the non-classicality of an optical channel can be experimentally certified even without having full tomographic knowledge of the process and, in particular, without assuming any specific model for the description of the gravitational interaction. Explicit experimental tests are described later.

Even if our proposal is model-independent, it is nevertheless useful to focus on a specific model of the dynamics. For this purpose, we explicitly evaluate the quantum steady-state of the system assuming that gravity can be approximated by a quadratic  Hamiltonian interaction (e.g.~a linearized Newtonian force) between the mechanical degrees of freedom of the two optomechanical systems. 
As discussed in Supplemental Material \cite{supplemental_material} \ref{app_commutators}, this direct Hamiltonian interaction can also be derived from the theory of linearized quantum gravity \cite{bose2022mechanism, christodoulou2023locally} whenever graviton emission is negligible. On the contrary, models in which gravity is described as a classical entity \cite{kafri2013noise, kafri2014classical, kafri2015bounds, tilloy2016sourcing, oppenheim2023postquantum, oppenheim2023gravitationally,tilloy2024general} are intrinsically irreversible \cite{galley2023any} and, therefore, incompatible with a bare Hamiltonian interaction.
In this setting and within the rotating-wave approximation, we show that gravity induces a quantum thermal attenuator channel \cite{serafini2023quantum, khatri2020principles} between independent optical frequency modes of the two subsystems. The GIT channel is completely characterized by two analytically computable parameters: the transmissivity $\eta$ and the effective thermal occupation number $N$. The non-classicality of thermal attenuators has been extensively studied within the context of quantum communication theory \cite{khatri2020principles}. 
For example, it is known that a quantum attenuator is not entanglement-breaking if and only if $\eta \ge (1 - \eta)N$ \cite{holevo2008entanglement, serafini2023quantum, kianvash2023low}. Moreover, in the asymptotic limit of many uses, the same non-classicality condition enables the perfect
transmission of quantum information (nonzero two-way quantum
capacity) \cite{mele2023maximum}.

Other protocols for testing the quantum nature of gravity
have been proposed in the literature \cite{mari2016experiments, marletto2017gravitationally, bose2017spin, belenchia2018quantum, danielson2022gravitationally, marshman2020locality, krisnanda2020observable, matsumura2020gravity, christodoulou2023locally, galley2022no,poddubny2024nonequilibrium,lami2024testing,toccacelo2025benchmarks, kryhin2025distinguishable, miao2020quantum,  datta2021signatures, angeli2025probing, carlesso2019testing, hanif2024testing} (see \cite{bose2025massive} and \cite{marletto2025quantum} for recent reviews). Most of them are based
on the generation of entanglement between two gravitationally coupled systems initially prepared in a separable state \cite{mari2016experiments, marletto2017gravitationally, bose2017spin, belenchia2018quantum, danielson2022gravitationally,  marshman2020locality, krisnanda2020observable, matsumura2020gravity, christodoulou2023locally, galley2022no, poddubny2024nonequilibrium, marletto2025quantum}.
Recently, it has been argued that the hypothesis of classical gravity could be falsified even without generating entanglement, by simply showing that the dynamics of gravitationally coupled mechanical resonators is inconsistent with a local classical description of the gravitational interaction  \cite{lami2024testing,toccacelo2025benchmarks, kryhin2025distinguishable, miao2020quantum, datta2021signatures, angeli2025probing, carlesso2019testing, hanif2024testing}. In the same spirit,
our work does not aim at generating gravitationally-induced
entanglement, but at assessing whether a gravitationally-induced quantum channel is non-classical (in the sense of non-entanglement-breaking). This is a weaker requirement than
observing entanglement generation, but still sufficient for ruling out local classical models of gravity. For example,
even in a regime in which the GIT channel is
non-classical, entanglement is typically absent in the stationary quantum state of the optomechanical systems.

\noindent {\it The model---} We consider two independent laser-driven optomechanical systems that we call $S_1$ and $S_2$ (see Fig.~\ref{fig:scheme}).
Each system $S_j$ is composed of an optical cavity mode of frequency $\omega_{A}$ driven by an external coherent laser pump of frequency $\omega_{\rm pump}$ and interacting with a mechanical resonator of frequency $\omega_B$ (e.g.~via radiation pressure). Assuming the system is driven into a stable semi-classical steady state and linearizing the dynamics around the classical mean-field solution, the effective optomechanical Hamiltonian of $S_j$ is \cite{milburn2011introduction, aspelmeyer2014cavity, barzanjeh2022optomechanics}
\begin{equation}\label{Hj}
H_j = \hbar \Delta a_j^\dag a_j + \hbar \omega_B b_j^\dag b_j + \hbar g (a_j^\dag + a_j) (b_j^\dag + b_j),
\end{equation}
where $a_j$ and $b_j$ are the annihilation operators of the optical and mechanical modes respectively, $\Delta=\omega_{A} - \omega_{\rm pump}$ is the optical detuning, and $g$ is the effective optomechanical coupling (proportional to the driving pump amplitude). We remark that this linearization is justified only in the strong–pump regime, in which the operators $a$ and $b$ represent small fluctuations around their respective mean fields \cite{GenesAdv, milburn2011introduction, aspelmeyer2014cavity, barzanjeh2022optomechanics}.
For simplicity, we use the same parameters for $H_1$ and $H_2$, but similar results also hold for asymmetric parameters, as discussed in Supplemental Material \cite{supplemental_material} \ref{app:asymmetric_case}. We assume that $S_1$ and $S_2$ are well isolated from each other, apart from a weak gravitational interaction between the mechanical modes that, for the moment, we avoid specifying to remain as general as possible. We remark that gravity must be the only interaction mediating the coupling between $S_1$ and $S_2$, meaning that, for example, any electromagnetic interaction is well screened (see e.g.~\cite{westphal2021measurement}).

We also take dissipation and thermal noise into account. We assume that the optical modes and the mechanical modes are characterized by damping rates $\kappa$ and $\gamma$ respectively and that they are coupled to their respective noise operators, $a_{{\rm in}_j}$ and $b_{{\rm in}_j}$ \cite{gardiner2004quantum}. The noise operators are optical and mechanical fields characterized by the following bosonic commutators:
\begin{equation}
[a_{{\rm in}_j}(t), a_{{\rm in}_j}(t')^\dag] = [b_{{\rm in}_j}(t), b_{{\rm in}_j}(t')^ \dag]=\delta(t-t'). \label{eq:commutators}
\end{equation}
For the mechanical modes, we assume a thermal Markovian environment with $\langle b_{{\rm in}_j} \rangle = 0$ and correlation functions
\begin{equation}
\langle b_{{\rm in}_j}(t)^\dagger b_{{\rm in}_j}(t')  \rangle = N_T \delta(t - t'), \; \langle b_{{\rm in}_j}(t) b_{{\rm in}_j}(t')  \rangle=0, \label{eq:mechanical_noise}
\end{equation}
where $N_T$ is the mean bosonic occupation number at temperature $T$ and mechanical frequency $\omega_B$, i.e., 
$N_T=[\exp(\frac{\hbar \omega_{\rm B}}{k_B T}) -1]^{-1}$, with $k_B$ denoting the Boltzmann constant.

The optical noise operators $a_{{\rm in}_j}(t)$ admit a clear physical interpretation since they correspond to the actual radiation fields entering the optical cavities from their unique input-output port. Such operators should be considered as weak fields in addition to the strong classical laser pumps (implicit in the mean-field linearization).
We assume the second cavity is not driven by any additional optical field beyond the laser pump. This means  $a_{{\rm in}_2}(t)$ represents vacuum noise with $\langle a_{{\rm in}_2}(t) \rangle =\langle a_{{\rm in}_2}(t)^\dag a_{{\rm in}_2}(t') \rangle =0$.
On the contrary, for the first optical cavity, we assume an arbitrary input probe field prepared in a generic quantum state, with the only restriction of being weaker than the classical pump. 
The motivation for assuming an arbitrary $a_{{\rm in}_1}(t)$ is that in this work we are interested in the gravity-induced quantum channel from the optical input of the first cavity to the optical output of the second cavity (see Fig.~\ref{fig:scheme}). For the same reason, we apply the input-output theory of a single-sided optical cavity \cite{gardiner2004quantum}, to obtain the output field emerging from the second cavity:
\begin{equation}
a_{{\rm out}_2}(t) = \sqrt{\kappa} a_2(t)  - a_{{\rm in}_2}(t). \label{eq:input-output}
\end{equation}

Without gravity, $S_1$ and $S_2$ would be completely independent and any signal encoded into $a_{{\rm in}_1}(t)$ would never be transmitted to $a_{{\rm out}_2}(t)$. With gravity, $S_1$ and $S_2$ are coupled, and part of the input signal may reach $a_{{\rm out}_2}(t)$. Thus gravity can activate an effective transmission line from $a_{{\rm in}_1}(t)$ to $a_{{\rm out}_2}(t)$, i.e., a quantum optical channel from $S_1$ to $S_2$ as shown in Fig.~\ref{fig:scheme}.
We call this phenomenon {\it gravitationally induced transparency} (GIT) because it is analogous to the electromagnetic \cite{fleischhauer2005electromagnetically} and optomechanical \cite{weis2010optomechanically} counterparts and because transparency is only induced in a narrow frequency window of the order of the mechanical effective linewidth (see Supplemental Material \cite{supplemental_material} \ref{app:linewidth}).

\noindent {\it Non-classicality criterion---} 
The key point of this work is relating the non-classicality of gravity to the non-classicality of the gravitationally induced optical channel from a mode of the input field $a_{{\rm in}_1}(t)$ to a mode of the output field $a_{{\rm out}_2}(t)$. Indeed, since gravity is the only mediator between $S_1$ and $S_2$, the experimental observation of a non-classical channel between the optical input of $S_1$ and the optical output of $S_2$ would imply that gravity cannot be described as a classical process such as, for example, a classical feedback-like mechanism ``reading'' the position of a particle and applying a related force on another particle \cite{kafri2013noise, kafri2014classical, kafri2015bounds, tilloy2016sourcing, oppenheim2023postquantum, oppenheim2023gravitationally,tilloy2024general}. Notice that this implication has been rigorously proved \cite{galley2022no, marletto2020witnessing,martin2023gravity} under the reasonable assumption that gravity acts as a local mediator. Moreover, within the specific framework of linearized quantum gravity, the non-classicality of the GIT channel can be directly linked to the non-commutative nature of the gravitational field, as discussed in Supplemental Material \cite{supplemental_material} \ref{app_commutators}.

A proper theory of gravity in the quantum limit is still unknown but, on the contrary, a quantum theory of channels between optical modes is scientifically mature and experimentally accessible \cite{serafini2023quantum, khatri2020principles}. In particular, given a channel $\Phi$ that produces an output quantum state $\rho_{\rm out}=\Phi(\rho_{\rm in})$ from an input quantum state $\rho_{\rm in}$, a standard classicality criterion is the {\it entanglement-breaking} property \cite{horodecki2003entanglement, holevo2008entanglement} (formally defined in the introduction of this work).
Thus, the associated non-classicality criterion is:
\begin{equation}\label{eq:criterion_A}
\text{Criterion A: } \Phi \text { is not entanglement-breaking}. \hspace{4em}
\end{equation}
While it is clear that this is a signature of non-classicality, one may wonder if other alternative criteria may be equally meaningful.
For example, other reasonable choices may be:  
\begin{align}
\text{Criterion B: }& \Phi \text { cannot be simulated by LOCC}, \label{eq:criterion_B} \\
\text{Criterion C: }& \Phi \text { has nonzero two-way quantum capacity,}\label{eq:criterion_C}
\end{align}
where LOCC stands for {\it local operations and classical communication}  \cite{khatri2020principles} and the two-way quantum capacity is the asymptotic rate at which perfect quantum information transmission can be achieved with many uses of $\Phi$ and two-way classical communication \cite{khatri2020principles}.

Each one of the mentioned criteria, if satisfied, would imply the quantum nature of gravity. Luckily, there is no need to check all of them since it is enough to only consider  Criterion A.
Indeed, it can be shown that Criteria A and B are equivalent. This is a consequence of the fact that any entanglement-breaking channel can be described as a classical measure-and-prepare operation \cite{horodecki2003entanglement} and, conversely, any channel implemented by only local operations and classical communication (LOCC) is intrinsically unable to distribute any entanglement. 
 The non-classicality Criterion C is instead strictly stronger than A, in the sense that $C \Rightarrow A$, but $A \nRightarrow C$ \cite{khatri2020principles}. This means that using $C$ to rule out classical gravity would be overkill since the weaker condition $A$ is sufficient for our scope. 

What we discussed is valid for any channel and, therefore, for any model in which the gravitational interaction acts as a mediator. An important special case is when the quantum channel acts independently on single frequency modes $\omega$ as a Gaussian phase-insensitive thermal attenuator $\Phi_\omega=\mathcal{E}_{\eta, N}$, characterized by a transmissivity $\eta(\omega)$ and an effective thermal occupation number $N(\omega)$ \cite{serafini2023quantum, kianvash2023low}. As we are going to show, this is indeed the case for a linearized Newtonian force, which can be seen as a limit of linearized quantum gravity \cite{bose2022mechanism, christodoulou2023locally}, but it may also be compatible with many alternative phenomenological models of gravity. For example, spontaneous-collapse models \cite{bassi2013models} or classical feedback-like models \cite{kafri2013noise, kafri2014classical, kafri2015bounds, tilloy2016sourcing, oppenheim2023postquantum, oppenheim2023gravitationally,tilloy2024general}, are similar to the Newtonian model with the addition of gravitational Gaussian noise acting on the dynamics of massive particles such that, in our setup, gravity would still induce a Gaussian channel.
For $\Phi=\mathcal{E}_{\eta, N}$, it can be shown \cite{holevo2008entanglement, mele2023maximum} that

\begin{equation}
A \Leftrightarrow B \Leftrightarrow C \Leftrightarrow  \frac{\eta(\omega)}{
\pq{1 - \eta(\omega)}
N(\omega)} > 1. \label{eq:criterion_attenuator}
\end{equation}

The above inequality provides a sharp non-classicality criterion for the gravity-induced thermal attenuator. If the ratio is smaller than 1, the channel is equivalent to a fully classical measure-and-prepare process. If the ratio is larger than 1, the channel enables not only entanglement distribution (A) but, in the limit of many uses, also the noiseless transmission of quantum information (C).

\noindent {\it Experimental protocols---} 
Assuming we have an experimental setup such that low-noise GIT is achieved, what is the actual experiment that should be performed to falsify the hypothesis of classical gravity?
We propose three explicit protocols ordered by increasing technical complexity and characterized by different underlying assumptions. 

The simplest protocol assumes that the gravity-induced channel has the structure of a phase-insensitive thermal attenuator with unknown transmissivity $\eta(\omega)$ and thermal noise $N(\omega)$. As we discussed, within the RWA, this assumption captures weak Newtonian gravity but also a large class of phenomenological alternative models of gravity.   Assuming $\Phi_\omega=\mathcal{E}_{\eta, N}$, one can characterize the gravity-induced channel in the same way in which an experimentalist would characterize an optical or microwave transmission line, i.e., injecting one or more coherent input signals to estimate the transmissivity $\eta(\omega)$ and the thermal background noise $
\pq{1 - \eta(\omega)}
N(\omega)$. From \eqref{eq:criterion_attenuator}, if the ratio of the two estimated quantities is larger than 1, classical gravity is falsified.

The second protocol is valid for a generic channel $\Phi$ and, therefore, is free from any assumption on the gravitational interaction and on the system dynamics. It is based on the same benchmarking protocol used more than 25 years ago to demonstrate the non-classicality of continuous-variable quantum teleportation in a model-independent way \cite{furusawa1998unconditional, 
zhang2003quantum}. The idea is to inject, at the frequency $\omega$ of maximum GIT, different coherent states $|\alpha_{\rm in}\rangle$ sampled from a Gaussian phase-space distribution corresponding to a thermal ensemble with mean photon number $N_{\rm in}$. For each input coherent state $|\alpha_{\rm in}\rangle$, one can measure the Q-function $Q_{\rm out}(\alpha)=\langle \alpha | \rho_{\rm out} | \alpha \rangle$ of the output state by heterodyne detection and, in particular,  the input-output fidelity $\mathcal F_{\alpha_{\rm in}}=Q_{\rm out}(\alpha_{\rm in})$. Averaging over different input coherent states, one can estimate the average input-output fidelity $\mathcal F = \overline{\mathcal F_{\alpha_{\rm in}}}$. The optimal average fidelity achievable by any classical strategy is $\mathcal F_{\rm classical}= (N_{\rm in} + 1) /(2 N_{\rm in} + 1)$\cite{hammerer2005quantum} which, for $N_{\rm in} \gg 1$, is approximately $1/2$ (see also \cite{lami2024testing, toccacelo2025benchmarks} for new different proofs). Thus, measuring $\mathcal F > \mathcal F_{\rm classical}$ would imply the non-classicality of the effective channel (Criteria $A$ and $B$) and, as a consequence, of the gravitational mediator.
A similar protocol, directly defined on mechanical modes instead of GIT  optical modes, has been recently proposed in \cite{toccacelo2025benchmarks} as a simplified version of the framework introduced in \cite{lami2024testing}.

Finally, the third protocol is technically more difficult but provides a stronger type of non-classicality, namely, gravitationally-induced entanglement distribution (i.e., the generation of entanglement between the output signal and an ancillary mode of the input signal). 
This can be achieved by sending through the GIT channel a mode of a two-mode-squeezed (i.e. entangled) optical field. If nonzero entanglement is detected between the output mode (transmitted through the channel) and the ancillary mode (unaffected by the channel), classical gravity is ruled out.

Note that sufficient conditions for nonzero entanglement can be obtained by measuring entanglement witnesses, i.e., the second moments of collective modes, independently of the underlying quantum state (Gaussianity is not required). So, this third protocol is model-independent like the second one.

\noindent {\it Explicit solution for a quadratic Hamiltonian interaction ---} 
We now assume an explicit model in which the gravitational interaction between 
the two mechanical resonators of $S_1$ and $S_2$ 
can be described by adding to the full Hamiltonian of the system a quadratic interaction term
\begin{equation}\label{eq:interaction}
V= \hbar \lambda (b_1^\dag + b_1)(b_2^\dag + b_2),
\end{equation}
where $\lambda$ represents the gravitational coupling rate. 
This model is a good approximation for a linearized Newtonian force, where the specific value of $\lambda$ originates from the Taylor expansion of the gravitational potential energy of the mechanical modes and depends on the specific geometric configuration. For example, as discussed in Supplemental Material \cite{supplemental_material} \ref{app:analysis_of_technical_requirements}, $\lambda = G m/ ( d^3 \omega_B)$ for two spheres of mass $m$ whose centers are at an average distance $d$. 

The full Hamiltonian $H = H_1 + H_2 + V$ can be further simplified in the so-called {\it rotating-wave approximation} (RWA). In this work, we are interested in enhancing  state-transfer interactions \cite{milburn2011introduction, aspelmeyer2014cavity}, so we focus on the resonance condition $\Delta=\omega_B$ and move to an interaction picture with respect to $H_0 = \hbar \omega_B (a_j^\dag a_j + b_j^\dag b_j)$ such that, after neglecting all terms rotating at frequency $\pm 2 \omega_B$ (RWA), we are left with a passive interaction Hamiltonian:
\begin{equation}
H_I = \hbar g (a_1 b_1^\dag +a_1^\dag b_1 + a_2 b_2^\dag +a_2^\dag b_2) + \hbar \lambda (b_1 b_2^\dag + b_2^\dag b_1).
\end{equation}
The above Hamiltonian represents a chain of 4 bosonic modes in which excitations (and therefore quantum information) can coherently hop from one mode to a nearest-neighbor in the chain.
Taking into account dissipation and thermal noise, we obtain the following set of quantum Langevin equations \cite{gardiner2004quantum, milburn2011introduction, aspelmeyer2014cavity}:
\begin{align}
\frac{d a_j(t)}{dt} &= - \frac{\kappa}{2} a_j(t) - i g  b_j(t) + \sqrt{\kappa} a_{{\rm in}_j}(t), \label{eq:langevin1} \\
\frac{d b_j(t)}{dt} &= - \frac{\gamma}{2} b_j(t) - i g a_j(t) - i \lambda b_{3-j}(t) + \sqrt{\gamma} b_{{\rm in}_j}(t), \nonumber
\end{align}
where $j\in \{1,2\}$, $\kappa$ and $\gamma$ are the optical and mechanical damping rates respectively, $a_{{\rm in}_j}$ and $b_{{\rm in}_j}$ are the optical and mechanical noise operators already introduced in Eqs.~(\ref{eq:commutators}) and (\ref{eq:mechanical_noise}).

We move to the frequency domain where the Langevin equations (\ref{eq:langevin1}) can be easily solved by linear algebra (see Supplemental Material \cite{supplemental_material}  \ref{app:full_solution}). Combining the solution of (\ref{eq:langevin1}) with Eq.~\eqref{eq:input-output}, we can express the output operator of the second cavity as a linear combination of the four input operators. In particular, identifying the term proportional to $a_{{\rm in}_1}(\omega)$ as the input ``signal'',  the GIT channel from  $a_{{\rm in}_1}(\omega)$  to $a_{{\rm out}_2}(\omega)$ can be written as a phase-insensitive thermal attenuator \cite{serafini2023quantum, kianvash2023low},
\begin{equation}
a_{{\rm out}_2}(\omega) =  \sqrt{\eta(\omega)} e^{i \varphi(\omega)}  a_{{\rm in}_1}(\omega) + \sqrt{1 - \eta(\omega)} a_E(\omega), \label{eq:attenuator}
\end{equation}
where the transmissivity $\eta(\omega)$ and the phase factor $\varphi(\omega)$ can be analytically computed (see Supplemental Material \cite{supplemental_material} \ref{app:full_solution} for explicit expressions).  In  Eq.~\eqref{eq:attenuator}, we collected the noise terms proportional to $b_{{\rm in}_1}(\omega)$, $b_{{\rm in}_2}(\omega)$, and $a_{{\rm in}_2}(\omega)$, into an single  environmental mode  $a_E(\omega)$.
The quantum state of $a_E(\omega)$ is a thermal state with $\langle a_E(\omega)\rangle =0$ and $\langle [a_E(\omega)]^\dag a_E(\omega')\rangle =N(\omega)\delta(\omega - \omega')$, where the mean occupation number $N(\omega)$ can be analytically computed and represents the effective thermal noise of the attenuator channel. 
The non-classicality of the quantum channel is completely determined by $\eta(\omega)$ and $N(\omega)$ and is independent from the phase factor $e^{i\varphi(\omega)}$.

\begin{figure}[!t]  
    \includegraphics[width=\linewidth]{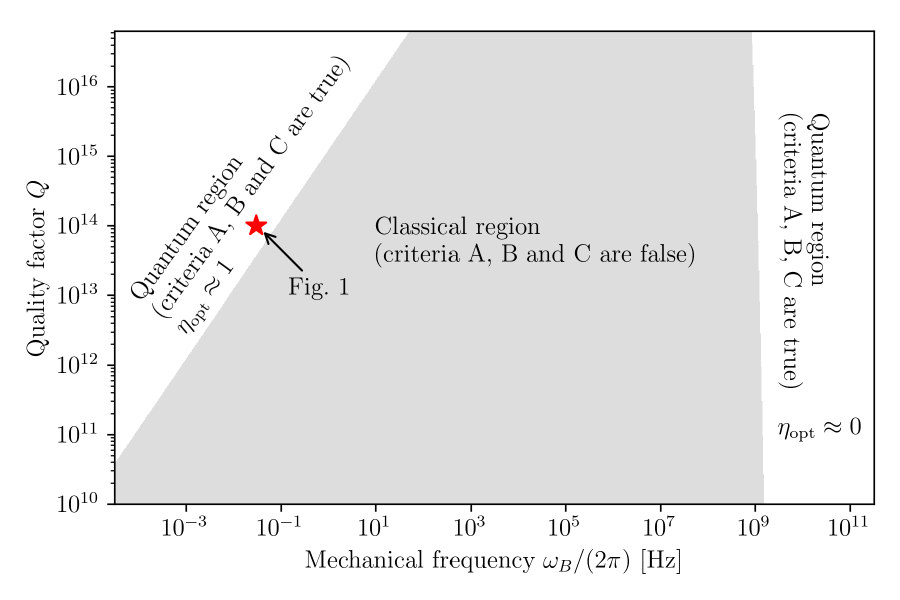}
    \caption{Non-classicality analysis of the GIT channel in the parameter space  $(\omega_B, Q=\omega_B/\gamma)$, according to Eq.~\eqref{eq:optimal_ratio} and Eqs.~(\ref{eq:criterion_A}-\ref{eq:criterion_C}). We assume the mechanical degrees of freedom are two nearby gold spheres at a temperature of $T=1$ mK, such that $\lambda \simeq \pi G \rho_{\rm Au}/(6\omega_B)$, where $\rho_{\rm Au}$ is the gold mass density. See Supplemental Material \cite{supplemental_material} \ref{app:analysis_of_technical_requirements} for more details. 
    The transmissivity $\eta$ of the effective channel is $\approx 1$  in the low-frequency region, but it is extremely small ($\lesssim 10^{-23}$) in the high-frequency region. This fact makes the high-frequency region theoretically valid but experimentally problematic. The red star corresponds to the parameters of Fig.~\ref{fig:scheme}.}
    \label{fig:parameter-space-main-text}
\end{figure}

Replacing the analytic solutions for $\eta(\omega)$ and $N(\omega)$ in Eq.~\eqref{eq:criterion_attenuator} and optimizing over $\omega$ and $g$ (see Supplemental Material \cite{supplemental_material} \ref{app:full_solution} for explicit calculations), we find that the maximum  ratio is achieved at $\omega_{\rm opt}=0$
\footnote{
Note that, in this picture, $\omega=0$ corresponds to the cavity resonance frequency $\omega_A$.}
and $g_{\rm opt}:=(\sqrt{\kappa}/2)(\gamma^2 +  4 \lambda^2)^{1/4}$, where Eq.~\eqref{eq:criterion_attenuator} reduces to
\begin{equation}
A \Leftrightarrow B \Leftrightarrow C \Leftrightarrow  \lambda^2 > \gamma^2 N_T (N_T + 1), \label{eq:optimal_ratio}
\end{equation}
corresponding to a thermal attenuator with:
\begin{equation}\label{eq:eta_opt_N_opt}
\eta_{\rm opt}= \frac{2 \lambda^2/\gamma^2}{1 + \sqrt{1 + 4 \lambda^2/\gamma^2}+ 2 \lambda^2/\gamma^2} , \quad N_{\rm opt}= N_T.
\end{equation}
We observe that, within the range of validity of the RWA (see App.~\ref{app:RWA}), the optimized gravity-induced channel is independent of the cavity decay rate $\kappa$, and its effective temperature reduces to the real temperature of the mechanical resonators. 
The analytic expression in Eq.~(\ref{eq:optimal_ratio}) determines a simple requirement for testing the non-classicality of gravity in terms of only three parameters ($\lambda, \gamma $ and $N_T$), providing a clear figure of merit in the search for the optimal parameter regions as shown in Fig.~\ref{fig:parameter-space-main-text}.
In principle, the condition in Eq.~(\ref{eq:optimal_ratio}) is the only fundamental requirement imposed by quantum mechanics. However, in a real experiment, one must consider further technical constraints such as ensuring reasonable values for the time duration of the experiment and for the magnitude of the effective transmissivity $\eta_{\rm opt}$. More details on technical requirements are given in Supplemental Material \cite{supplemental_material} \ref{app:analysis_of_technical_requirements}.

\noindent {\it Conclusions---} 
We introduced the concept of GIT, i.e., an effective optical transmission line induced by gravity acting as a mediator between two independent optomechanical systems. We proposed model-independent protocols for indirectly assessing the non-classicality of the gravitational interaction by testing the non-classicality of the GIT channel.
For the specific model of a quadratic gravitational potential, the gravity-induced channel reduces to a thermal attenuator that can be completely characterized analytically.

By framing the question about the quantum nature of gravity as a non-classicality test for a quantum optical channel, the problem is significantly simplified both conceptually and experimentally by using standard optical sources and detectors (see \ref{App_GITvsGIE} in Supplemental Material \cite{supplemental_material}). However, the experimental implementation of the approach proposed in this work remains very challenging. The reason is that, to our knowledge, all existing optomechanical devices operate in what we called "classical region" of the parameter space (see Fig.~\ref{fig:parameter-space-main-text}). 
Testing the GIT channel in the quantum regime requires developing new extraordinary devices with low mechanical frequencies and ultra-high quality factors. The remarkable history of gravitational-wave interferometers demonstrates that similar technological endeavors are possible.

In the short term, it would still be valuable to experimentally demonstrate even a classical instance of the GIT phenomenon for several reasons. First, it would represent an important technological milestone toward the ideal quantum regime. Second, classical GIT may be of interest also within the context of classical weak-gravity experiments at the microscopic scale \cite{westphal2021measurement, lee2020new}. Third, even in the classical regime, one might indirectly extrapolate the ideal results achievable in the quantum regime. For example, one could evaluate the input-output average fidelity $\mathcal F$ of the GIT channel (see Section {\it Experimental protocols}) at different temperatures and extrapolate an estimate of $\mathcal F$ at $T=0$. An extrapolated value larger than $\mathcal F_{\rm classical}$  would provide indirect evidence for the quantum nature of gravity, even if not as compelling as a direct experiment in the quantum regime.
Additionally, a different type of short-term experiment may be obtained by replacing the gravitational interaction with the Coulomb interaction between two charged mechanical degrees of freedom \cite{poddubny2024nonequilibrium}. Demonstrating a non-classical optical channel induced by a Coulomb interaction would constitute a proof-of-concept milestone towards the ideal gravitational experiment.

\noindent {\it Acknowledgements---} 
This research was supported by the MUR PNRR Extended Partnership NQSTI - PE00000023 - CUP J13C22000680006.

\noindent {\it Data availability---}
Code reproducing the results of this work can be found at  \cite{code}.

\bibliography{refs2}
\bibliographystyle{unsrturl}
\clearpage
\onecolumngrid

\begin{center}
{\bf\large  Supplemental material for ``Can gravity mediate the transmission of quantum information?''
}

\bigskip 
{Andrea Mari$^1$, Stefano Zippilli$^1$, David Vitali$^{1,2,3}$}

\vspace{0.1cm}
{\it \small
$^1$ Physics Division, School of Science and Technology, Universit\`a di Camerino, 62032 Camerino, Italy\\
}
{\it \small
$^2$ INFN, Sezione di Perugia, I-06123 Perugia, Italy\\
}
{\it \small
$^3$ CNR-INO, L.go Enrico Fermi 6, I-50125 Firenze, Italy\\
}
\end{center}


\renewcommand{\thesection}{S.\Roman{section}}
\renewcommand{\theequation}{S.\arabic{equation}}
\renewcommand{\thefigure}{S.\arabic{figure}}

\setcounter{section}{0}
\setcounter{equation}{0}
\setcounter{figure}{0}
\setcounter{page}{1}


\section{Analytic solution assuming a quadratic gravitational interaction potential} \label{app:full_solution}

Our starting point is the system of quantum Langevin equations \eqref{eq:langevin1} introduced in the main text.
Defining the array of annihilation operators $\mathbf{r}(t)=(a_1(t), b_1(t), a_2(t), b_2(t))^\top$ and the array of noise operators $\mathbf{w}(t)=(a_{\rm in_1}(t), b_{\rm in_1}(t), a_{\rm in_2}(t), b_{\rm in_2}(t))^\top$, the Langevin equations \eqref{eq:langevin1} can be written in matrix form as:
\begin{equation}
{\mathrm d}{\mathbf r}(t)/{\mathrm dt} = A {\mathbf r}(t) + B {\mathbf w}(t), \quad 
A = - \begin{pmatrix}
\frac{\kappa}{2} & i g & 0 & 0\\
i g & \frac{\gamma}{2} & 0 & i \lambda \\
0& 0 &  \frac{\kappa}{2} & i g \\
0& i \lambda &  i g  & \frac{\gamma}{2}
\end{pmatrix},
\quad B = \text{diag}(\sqrt{\kappa}, \sqrt{\gamma},\sqrt{\kappa},\sqrt{\gamma}).
\end{equation}
We now apply the Fourier transform $O(\omega)= \frac{1}{\sqrt{2 \pi}} \int_{-\infty}^{\infty} dt e^{i \omega t} O(t)$ to all bosonic operators.
In the frequency domain, the differential equation becomes $-i \omega {\mathbf r}(\omega) = A {\mathbf r}(\omega)+ B {\mathbf w}(\omega)$, whose solution is:
\begin{equation}
{\mathbf r}(\omega) = - (A + i \omega)^{-1}B   {\mathbf w}(\omega).
\end{equation}
In particular, the third row of the above equation can be written as:
\begin{equation}
a_2(\omega) = \tilde \alpha_1(\omega) a_{\rm in_1}(\omega) + \tilde \beta_1(\omega) b_{\rm in_1}(\omega) + \tilde \alpha_2(\omega) a_{\rm in_2}(\omega) + \tilde \beta_2(\omega) b_{\rm in_2}(\omega),
\end{equation}
where 
\begin{align}
\tilde \alpha_1(\omega) &= \sqrt{\kappa} \frac{i g^2  \lambda}{\det{A + i \omega}}\ , \\
\tilde \beta_1(\omega) &= \sqrt{\gamma}g  \frac{
\lambda (i \omega - \kappa/2)}{\det{A + i \omega}} \ , \\
\tilde \alpha_2(\omega) &=\sqrt{\kappa} \frac{g^2 \gamma/2 - i g^2 \omega + \kappa \gamma^2/8 - i \kappa \gamma \omega/2 + \kappa \lambda^2/2  - \kappa \omega^2/2 - i \gamma^2 \omega /4 - \gamma \omega^2 - i \lambda^2 \omega + i \omega^3
}{\det{A + i \omega}} \ , \\
\tilde \beta_2(\omega) &=\sqrt{\gamma} g \frac{- i g^2 - i \kappa \gamma / 4 - \kappa \omega/2 - \gamma \omega/2 + i \omega^2 }{\det{A + i \omega}} \ .
\end{align}
Using the input-output relation \eqref{eq:input-output}, we obtain the output field of the second cavity
\begin{equation}\label{eq:second_cavity_output}
a_{\rm out_2}(\omega) = \alpha_1(\omega) a_{\rm in_1}(\omega) + \beta_1(\omega) b_{\rm in_1}(\omega) + \alpha_2(\omega) a_{\rm in_2}(\omega) +  \beta_2(\omega) b_{\rm in_2}(\omega),
\end{equation}
where
\begin{equation}\label{coeff}
\alpha_1(\omega) = \sqrt{\kappa} \tilde \alpha_1(\omega), \quad \beta_1(\omega) = \sqrt{\kappa} \tilde \beta_1(\omega), \quad \alpha_2(\omega) = \sqrt{\kappa} \tilde \alpha_2(\omega) - 1, \quad \beta_2(\omega) = \sqrt{\kappa} \tilde \beta_2(\omega).
\end{equation}
One can check that $\sum_i |\alpha_i(\omega)|^2 + |\beta_i(\omega)|^2=1$, ensuring $a_{\rm out_2}(\omega)$ satisfy the bosonic commutation rules as expected. 

Now, we identify the term $\alpha_1(\omega) a_{{\rm in}_1}(\omega)$ in Eq.~\eqref{eq:second_cavity_output} as the input ``signal''  and we cast the remaining three terms into a single effective environmental mode:

\begin{equation}\label{eq:env_mode}
a_E(\omega):=  \frac{\beta_1(\omega)  b_{{\rm in}_1}(\omega)+ 
\alpha_2(\omega) a_{{\rm in}_2}(\omega) + \beta_2(\omega)  b_{{\rm in}_2}(\omega)}{\sqrt{1- |\alpha_1(\omega)|^2}}.
\end{equation}
The normalization factor in the denominator of the above equation ensures $a_E(\omega)$ obeys the canonical commutation rules of a bosonic mode. Its quantum state is a thermal state with  $\langle a_E(\omega)\rangle =0$ and $\langle [a_E(\omega)]^\dag a_E(\omega')\rangle =N(\omega)\delta(\omega - \omega')$, which is completely characterized by the effective mean occupation number:
\begin{equation}
N(\omega) = \frac{|\beta_1(\omega)|^2+|\beta_2(\omega)|^2}{1 - |\alpha_1(\omega)|^2}N_T. \label{eq:N}
\end{equation}
Rewriting Eq.~\eqref{eq:second_cavity_output} as 
\begin{equation}
a_{{\rm out}_2}(\omega) =  \sqrt{\eta(\omega)} e^{i \varphi(\omega)}  a_{{\rm in}_1}(\omega) + \sqrt{1 - \eta(\omega)} a_E(\omega), \label{eq:attenuator_app}
\end{equation}
we recognize a familiar structure: the GIT channel from the optical input of the first cavity  $a_{{\rm in}_1}(\omega)$  to the optical output of the second cavity $a_{{\rm out}_2}(\omega)$ is a phase-insensitive thermal attenuator \cite{serafini2023quantum, kianvash2023low} with transmissivity $\eta(\omega)=|\alpha_1(\omega)|^2$ and thermal occupation number $N(\omega)$ given in Eq.~\eqref{eq:N}. Notice that the phase factor  $e^{i\varphi(\omega)}=\alpha_1(\omega)/|\alpha_1(\omega)|$ is irrelevant for the quantum properties of the channel since it can be absorbed into a redefinition of the input operator $a_{{\rm in}_1}'(\omega)= e^{i\varphi(\omega)} a_{{\rm in}_1}(\omega)  $.
The explicit expression of the effective transmissivity of the GIT channel $a_{\rm in_1}(\omega) \rightarrow a_{\rm out_2}(\omega)$ is:
\begin{equation}
\eta(\omega) = |\alpha_1(\omega)|^2=\frac{k^2 g^4 \lambda^2}{|\det{A + i \omega}|^2}.
\end{equation}
The explicit expression for the effective thermal occupation number of the GIT channel is:
\begin{equation}\label{eq:N_appendix}
N(\omega) =\frac{|\beta_1(\omega)|^2+|\beta_2(\omega)|^2}{1 - |\alpha_1(\omega)|^2}N_T = \kappa \gamma g^2  N_T \frac{\lambda^2(\kappa^2/4 +\omega^2) + (g^2 + \kappa \gamma/4 - \omega^2)^2 + (\kappa \omega/2 + \gamma \omega/2)^2 }{(1 - \eta(\omega))|\det{A + i \omega}|^2}.
\end{equation}

When evaluating the non-classicality criterion introduced in Eq.~\eqref{eq:criterion_attenuator} of the main text, the term $|\det(A + i \omega)|^2$ simplifies and we obtain
\begin{equation}
\frac{\eta(\omega)}{[1 - \eta(\omega)]N(\omega)}= 
\frac{|\alpha_1(\omega)|^2}{N_T[|\beta_1(\omega)|^2+|\beta_2(\omega)|^2]}
= \frac{g^2 \kappa \lambda^2}
{N_T \gamma [\lambda^2(\kappa^2/4 +\omega^2) + (g^2 + \kappa \gamma/4 - \omega^2)^2 + (\kappa \omega/2 + \gamma \omega/2)^2 ]} \ge 1. \label{eq:general_ratio}
\end{equation}

\begin{figure}[!h] 
    \includegraphics[width=0.7\linewidth]{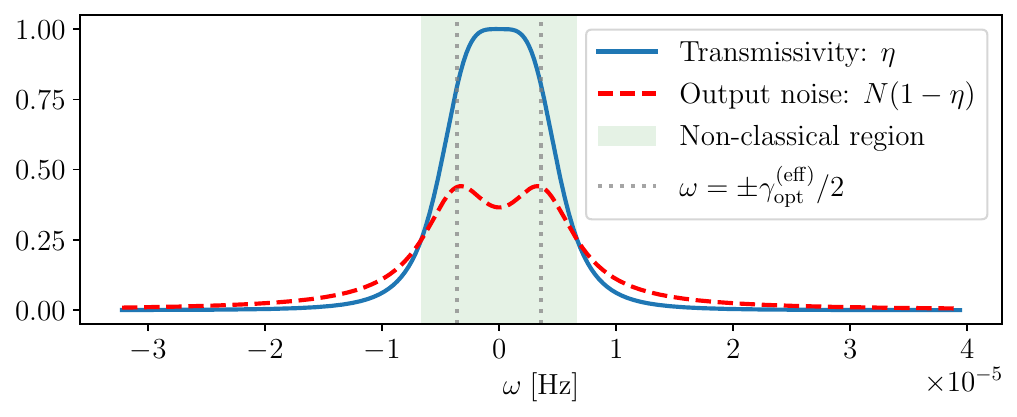}
    \caption{Spectral analysis of the gravitationally-induced transparency (GIT) phenomenon. The transmissivity $\eta(\omega)$ (full line) and the output noise $[1 - \eta(\omega)] N(\omega)$ (dashed line) for different frequencies of the input mode $a_{\rm in_1}(\omega)$. The frequency is defined in interaction picture, such that the origin $\omega=0$ is centered at the cavity frequency $\omega_A$. The frequency window between the two dotted lines is the estimated transparency linewidth according to Eq.~\eqref{eq:linewidth}. The light-green region highlights the frequency window in which the effective transmission channel $\Phi_\omega=\mathcal{E}_{\eta, N}$ is a non-classical quantum attenuator according to \eqref{eq:general_ratio}, i.e., where the transmissivity exceeds the output noise. The physical parameters used in this plot are:
    $\omega_B=2 \pi\times0.03$ Hz, $\gamma=10^{-14} \omega_B$, $\kappa=0.1 \omega_B$, $\lambda= w_G^2/\omega_B=3.58\times 10^{-6}$ Hz, $N_T=[\exp(\omega_B/w_T) - 1]^{-1}=6.94 \times 10^8$, where $w_G$ is given by Eq.~\eqref{eq:grav_w} assuming two nearby gold spheres and where $w_T$ is given by \eqref{eq:grav_w} assuming a temperature of $T=1$ mK. The optomechanical coupling rate is set to its optimal value $g_{\rm opt}=(\sqrt{\kappa}/2)(\gamma^2 +  4 \lambda^2)^{1/4}\simeq 1.84 \times 10^{-4}$ Hz, according to Eq.~\eqref{eq:optimal_parameters}.}
    \label{fig:git}
\end{figure}

In Fig.~\ref{fig:git}, we plot the transmissivity $\eta(\omega)$ and the output noise $[1 - \eta(\omega)]N(\omega)$ for a specific choice of parameters. The plot demonstrates two main phenomena: the emergence of GIT in a narrow frequency band at the cavity resonance frequency and its non-classical nature according to Eq.~\eqref{eq:general_ratio}.

\subsection{The special case $\omega=0$}\label{App_symm_res}
We expect the optimal transmission to happen at the resonance frequency $\omega_A$ of the optical cavities which, in the reference frame used for performing the rotating wave approximation, corresponds to $\omega=0$. So we first focus on this special case. Later, we will show that the non-resonant case $\omega \neq 0$ is indeed suboptimal.
Evaluating \eqref{eq:general_ratio} at $\omega=0$ and maximizing it  over $g$, we get
\begin{equation}\label{eq:optimal_parameters}
g_{\rm opt}=\frac{\sqrt{\kappa}(\gamma^2 +  4 \lambda^2)^{1/4}}{2} , \quad \left.
\frac{\eta(\omega)}{N(\omega) (1 - \eta(\omega))}\right \vert_{\omega=0, \; g=g_{\rm opt}}= \frac{2 \lambda^2/\gamma^2 }
{N_T \left(1 + \sqrt{1 + 4 \lambda^2/\gamma^2}\right)} \ge 1  \Longleftrightarrow   \lambda^2 > \gamma^2 N_T (N_T + 1),
\end{equation}
which is the non-classicality condition reported in the main text in Eq.~\eqref{eq:optimal_ratio}.
The corresponding optimal transmissivity is:

\begin{equation}
\eta_{\rm opt}=\left.\eta(\omega)\right \vert_{\omega=0, \; g=g_{\rm opt}}= \frac{2 \lambda^2/\gamma^2}{1 + \sqrt{1 + 4 \lambda^2/\gamma^2}+ 2 \lambda^2/\gamma^2} \simeq
\begin{cases}
1, & \text{for } \gamma \ll \lambda, \\
\\
\lambda^2/\gamma^2,  & \text{for } \gamma \gg \lambda. \\
\end{cases}
\end{equation}
meaning that for a small mechanical damping rate $\gamma \ll \lambda$, a near-perfect gravitationally-induced transparency is achievable at $\omega=0$. In the opposite regime of a large mechanical damping rate $\gamma \gg \lambda$, the maximum transmissivity is of the order of $\lambda^2 / \gamma^2 \ll 1$, making the experimental detection of non-classicality very difficult due to the strong signal attenuation.
We can also explicitly evaluate the optimal effective thermal occupation number $N(\omega)$ defined in Eq.~\eqref{eq:N_appendix}, obtaining the simple identity: 
\begin{equation}
N_{\rm opt}=\left. N(\omega) \right \vert_{\omega=0, \; g=g_{\rm opt}} = N_T,
\end{equation}
meaning that the effective temperature of the optimal gravity-induced attenuator corresponds to the real temperature of the mechanical resonators.
A further interesting observation is the following:
\begin{equation}\label{eq:zero_reflection}
\left. \alpha_2 \right \vert_{\omega=0, \; g=g_{\rm opt}} = 0,
\end{equation}
where $\alpha_2$ is the coefficient appearing in Eq.~\eqref{eq:second_cavity_output}
The above equation has a quite intuitive physical interpretation: the optimal gravity-induced transparency is obtained at zero optical reflection. Indeed, taking into account the symmetry between the exchange of $S_1$ and $S_2$, Eq.~\eqref{eq:zero_reflection} implies that any signal encoded in $a_{\rm in_j}$ does not leak from the corresponding cavity output $a_{\rm out_j}$ such that the transmission to the opposite cavity output $a_{\rm out_{3-j}}$ is maximized.

\subsection{Suboptimality of the $\omega\neq 0$ case}

Maximizing the ratio in Eq.~\eqref{eq:general_ratio} with respect to $\omega$ we get three critical points:
$\omega=0$ and $\omega = \pm \omega'$, where $\omega'=\pm \sqrt{g^2 - \kappa^2/8 - \gamma^2/8- \lambda^2/2}$. We already considered the special case $\omega=0$. Moreover, since the quantity we are optimizing is even with respect to $\omega$, we can focus on just $\omega = \omega'$. In this case, the ratio is a monotonically increasing function of $g$ and we have
\begin{equation}
\left.\frac{\eta}{N (1 - \eta)}\right \vert_{\omega=\omega'} < \left.\lim_{g \rightarrow \infty} \frac{\eta}{N (1 - \eta)}\right \vert_{\omega=\omega'}= \frac{\kappa \lambda^2}{N_T \gamma (4\kappa^2 + \kappa\gamma/2 + \gamma^2/4 + \lambda^2)}\ .
\end{equation}
Optimizing the right-hand-side over $\kappa$, we get $\kappa'=\sqrt{\gamma^2+4 \lambda^2}$ and
\begin{equation}
\left.\frac{\eta}{N (1 - \eta)}\right \vert_{\omega=\omega'} < 
\left.\lim_{g \rightarrow \infty} \frac{\eta}{N (1 - \eta)}\right \vert_{\omega=\omega', \kappa=\kappa'}=
 \frac{2 \lambda^2/\gamma^2 }
{N_T \left(1 + \sqrt{1 + 4 \lambda^2/\gamma^2}\right)}
=  \left.
\frac{\eta}{N (1 - \eta)}\right \vert_{\omega=0, \; g=g_{\rm opt}}.
\end{equation}
The above expression shows that the non-resonant case $\omega\neq 0$ is always suboptimal with respect to the resonant case $\omega=0$ considered in Eq.~\eqref{eq:optimal_parameters}.

\subsection{Transparency linewidth of the gravity-induced channel}\label{app:linewidth}
We have just shown that the optimal gravity-induced transmission is achieved at $\omega=0$ (relative to the cavity frequency $\omega_A$), but how large is the frequency band around $\omega=0$ in which we have a significant transparency?
Since the GIT channel is mediated by the mechanical resonators and since the mechanical resonators have a narrow frequency response, we expect a narrow transparency window of the order of the effective linewidth of the mechanical resonators.

Therefore, to estimate the transparency frequency window, we need to estimate the effective damping rate of the mechanical resonators.
The laser detuning considered in this work is $\Delta=\omega_{\rm B}$, corresponding to the so-called laser-cooling regime that has been extensively studied in the field of quantum optomechanics \cite{milburn2011introduction, aspelmeyer2014cavity}.
In this laser-cooling regime, and within the RWA, the mechanical damping rate $\gamma$ is increased by the optical cavity mode such that the effective damping rate is $\gamma^{\rm (eff)} \simeq \gamma + 4 g^2/\kappa$ \cite{aspelmeyer2014cavity}.
We have already shown that the gravitationally induced transmission is optimized for $g_{\rm opt}=(\sqrt{\kappa}/2)(\gamma^2 +  4 \lambda^2)^{1/4}$ (see Eq. \eqref{eq:optimal_parameters}), yielding the following estimate of the transparency linewidth: 
\begin{equation} \label{eq:linewidth}
\gamma^{\rm (eff)}_{\rm opt} \simeq \gamma + g_{\rm opt}^2/\kappa  = \gamma + \sqrt{\gamma^2 + 4 \lambda^2}
\simeq
\begin{cases}
2 \lambda , & \text{for } \gamma \ll  \lambda, \\
\\
2 \gamma,  & \text{for } \gamma \gg \lambda. \\
\end{cases}
\end{equation}

\subsection{Considerations regarding the rotating wave approximation}
\label{app:RWA}

The model analyzed in this work is based on the rotating wave approximation, in which non-resonant processes are neglected. For this approximation to be valid, the rates of dissipative processes must be much smaller than the relevant system frequencies.  

Focusing on a single optomechanical system (disregarding, for the moment, the gravitational interaction), this approximation amounts to neglecting the effect of back-action noise. In the weak coupling limit $g \ll \kappa$ (consistent with the parameters of Fig.~\ref{fig:scheme} in the main text and Fig.~\ref{fig:git}), this is valid if the natural thermal mechanical noise rate $\gamma\ N_T$ is much larger than the light-induced mechanical noise rate $\gamma\oo\ N\oo$, where $\gamma\oo$ is the light-induced dissipation rate, and $N\oo$ describes the effective number of thermal mechanical excitations associated with residual heating due to light scattering. Under the resonant condition analyzed in this work, $\gamma\oo = 4g^2/\kappa$, and $N\oo$ is given by the sideband cooling limit, $N\oo = \kappa^2 / 4\omega_B^2$. This implies that for our model to be valid, it is necessary that  
$N_{T}\gg\frac{g^2}{ \gamma}\frac{\kappa}{\omega_B^2}$. This condition is fulfilled for the parameters in Fig.~\ref{fig:scheme}.  

We note that in the low mechanical frequency regime, the validity of this approximation requires particularly low cavity linewidths $\kappa$. This is one of the challenging aspects of our proposal. 
For larger values of $\kappa$ one has to relax the rotating wave approximation, so that the expression for the output spectral mode at frequency omega~\rp{eq:second_cavity_output} must be generalized to include the coupling with the input spectral modes at the frequency $-\omega-2\omega_B$, taking the form
\begin{eqnarray}
\label{eq:second_cavity_output_noRWA}
a_{\rm out_2}(\omega) &=& 
  \alpha_1(\omega)\ a_{\rm in_1}(\omega) 
+ \beta_1(\omega)\ b_{\rm in_1}(\omega) 
+ \alpha_2(\omega)\ a_{\rm in_2}(\omega) 
+ \beta_2(\omega)\ b_{\rm in_2}(\omega)
\nn\\&&
+ \mu(\omega)\ a_{\rm in_1}\da(\omega+2\omega_B) 
+ \nu_1(\omega)\ b_{\rm in_1}\da(\omega+2\omega_B) 
+ \mu_2(\omega)\ a_{\rm in_2}\da(\omega+2\omega_B) 
+ \nu_2(\omega)\ b_{\rm in_2}\da(\omega+2\omega_B),
\end{eqnarray}
where $a_{\rm in_1}\da(\omega+2\omega_B)=\pq{a_{\rm in_1}(-\omega-2\omega_B)}\da$.
This implies that our optical channel from $a_{{\rm in}_1}(\omega)$ to $a_{{\rm out}_2}(\omega)$ would couple to additional, uncontrolled dissipative ports through the input fields at $-\omega-2\omega_B$. As a result, additional noise would be introduced, reducing the
non-classicality of the single-mode channel.  
A more efficient test in this scenario would probably require analyzing the two-mode channel from the two input modes $a_{{\rm in}_1}(\omega)$ and $a_{{\rm in}_1}(-\omega-2\omega_B)$ to the two output modes $a_{{\rm out}_2}(\omega)$ and $a_{{\rm out}_2}(-\omega-2\omega_B)$. Notice that non-classicality criteria exist for general multi-mode Gaussian channels \cite{holevo2008entanglement}, but they are not as simple as the criterion for a single-mode attenuator introduced in \eqref{eq:criterion_attenuator} of the main text.

\section{The general case in which $S_1$ and $S_2$ are not symmetric.} \label{app:asymmetric_case}

In this work, we focused on the simple case in which the optomechanical systems $S_1$ and $S_2$ have the same parameters ($\omega_A$, $\omega_B$, $\kappa$, $\gamma$, $N_T$, etc.) because it significantly simplifies the analytic derivation of the main results. However, in this Section, we show that the symmetry between $S_1$ and $S_2$ is not a strong assumption and that similar results can be obtained even in a regime in which $S_1$ and $S_2$ are asymmetric. 
Specifically, here we assume that the two systems can be different with mechanical frequencies $\omega_{B,j}$, mechanical dissipation rates $\gamma_j$, optomechanical couplings $g_j$, detunings $\Delta_j=\omega_{A,j}-\omega_{{\rm pump}, j}$, and optical linewidths $\kappa_j$, for $j\in{1,2}$. In any case, we assume that the parameters allow for the rotating wave approximation as discussed in the main text. In particular, we assume that the mechanical frequencies are sufficiently close $\abs{\omega_{B,1}-\omega_{B,2}}\ll\omega_{B,1},\omega_{B,2}$.
In this case, it is useful to use a different picture where the optical fields rotate at the laser frequencies [as in Eq.~\rp{Hj} of the main text]. The quantum Langevin equations in Fourier space are given by
\begin{eqnarray}\label{QLE_asymm}
-\pq{\frac{\kappa_j}{2}+\ii\pt{\Delta_j-\omega}}\ a_j(\omega)-\ii\ g_j\ b_j(\omega)+\sqrt{\kappa_j}\ a_{{\rm in}_j}(\omega)&=&0
\nn \ ,\\
-\pq{\frac{\gamma}{2}+\ii\pt{\omega_{B,j}-\omega}}\ b_j(\omega)+\ii\ g_j\ a_j(\omega)-\ii\ \lambda\ b_{3-j}(\omega)+\sqrt{\gamma_j}\ b_{{\rm in}_j}(\omega)&=&0\ .
\end{eqnarray}
In this picture the frequency $\omega=0$ corresponds the frequency of the laser fields, meaning that the input spectral mode at optical frequency $\omega_{\rm pump,1}+\omega$ is resonantly coupled to the output spectral mode at frequency $\omega_{\rm pump,2}+\omega$. 

Introducing the optomechanical and gravitational cooperatives  
\begin{eqnarray}
\Gamma_j&=&\frac{4\ g_j^2}{\kappa_j\ \gamma_j}\ , \qquad \Gamma_\lambda = \frac{4\ \lambda^2}{\gamma_1\ \gamma_2}\ ,
\end{eqnarray}
the relative detunings
\begin{eqnarray}
x_j(\omega)&=&2\ \frac{\Delta_j-\omega}{\kappa_j}\ , \qquad
y_j(\omega)=2\ \frac{\omega_{B,j}-\omega}{\gamma_j}\ ,
\end{eqnarray}
and the parameters
\begin{eqnarray}
\varrho(\omega)&=&\
\pq{1+\ii\ x_1(\omega)}\pq{1+\ii\ y_1(\omega)}+\Gamma_1\ ,
\qquad
\varsigma(\omega)=
\pq{1+\ii\ x_2(\omega)}
\pg{\pq{1+\ii\ y2(\omega)}
+
\frac{\pq{1+\ii\ x_1(\omega)}\ \Gamma_\lambda}{\varrho(\omega)}
}\ ,
\end{eqnarray}
we find the following expressions for the coefficients~\rp{coeff}
\begin{eqnarray}
\alpha_1(\omega)&=&
2\ \ii\ \frac{\sqrt{\Gamma_\lambda\ \Gamma_1}}{\varrho(\omega)}\ \frac{\sqrt{\Gamma_2}
}{
\varsigma(\omega)+\Gamma_2 
}
\nn \ ,\\
\beta_1(\omega)&=&
-\frac{2\ 
(1+\ii\ x_1)\ \sqrt{\Gamma_\lambda}}{\varrho(\omega)}\
\frac{\sqrt{\Gamma_2}
}{
\varsigma(\omega)+\Gamma_2 
}
\nn \ ,\\
\beta_2(\omega)&=&
-2\ \ii\
\frac{\sqrt{\Gamma_2}
}{
\varsigma(\omega)+\Gamma_2
}\ .
\end{eqnarray}
Note that the resonant situation with symmetric systems analyzed in Sec.~\ref{App_symm_res} corresponds to the condition $x_j(\omega)=y_j(\omega)=0$, that is $\omega=\omega_{B,1}=\omega_{B,2}=\Delta_1=\Delta_2$.
In the asymmetric case, the criterion~\rp{eq:general_ratio} has to account for the fact that the two different mechanical resonators may have a different number of thermal excitations $N_{T,j}$ such that it takes the form 
\begin{eqnarray}
\frac{
\abs{\alpha_1(\omega)}^2
}{
\abs{\beta_1(\omega)}^2\ N_{T,1}+
\abs{\beta_2(\omega)}^2\ N_{T,2}
}>1\ ,
\end{eqnarray}  
and using the expressions identified above it reduces to the form
\begin{eqnarray}\label{criterion_asymm}
\frac{
\Gamma_1
}
{
\abs{1+\ii\ x_1(\omega)}^2\ N_{T,1}+\abs{\varrho(\omega)}^2\ \frac{N_{T,2}}{\Gamma_\lambda}
}>1\ .
\end{eqnarray}
Interestingly, this shows that the non-classicality criterion depends on the second system only through the mechanical thermal noise parameter $\gamma_2\ N_{T,2}$.
This allows for the optimization of this ratio and of the transmission coefficient $\eta$ also for asymmetric systems. 
We find that the maximum of the ratio on the left hand side of Eq.~\rp{criterion_asymm} is found for $\Gamma_1=\Gamma_{{\rm opt},1}(\omega)$ and $x_1(\omega)=x_{{\rm opt},1}(\omega)$ with
\begin{eqnarray}\label{Gammaopt1}
\Gamma_{{\rm opt},1}(\omega)&=&
\sqrt{
\pq{\Gamma_\lambda\ \frac{N_{T,1}}{N_{T,2}} +1+y_1^2(\omega)}
\pq{1+\frac{y_1^2(\omega)}{
\pt{1+2\ N_{T,1}}^2
}}
}
\ ,\\
x_{{\rm opt},1}(\omega)&=&\frac{y_1(\omega)}{1+2\ N_{T,1}}\ .
\end{eqnarray}
Under these conditions the criterion~\rp{criterion_asymm} reduces to
\begin{eqnarray}\label{criterion_asymm_opt}
\frac{\Gamma_\lambda}{4\ N_{T,2}\pt{1+N_{T,1}}}>1\ ,
\end{eqnarray}
which is equivalent, in the symmetric case, to Eq.~\rp{eq:optimal_ratio} of the main text.
This result shows that the same maximum of the ratio in the criterion~\rp{criterion_asymm} can be obtained for any spectral mode by properly tuning the amplitude and frequency of the pump laser of the first system.
However, it is possible to show that the maximum of the transmission parameter 
$\eta(\omega)=\abs{\alpha_1(\omega)}^2$ expressed by Eq.~\rp{eq:eta_opt_N_opt} of the main text can be obtained only for 
$y_1(\omega)=0$ 
or, equivalently, for the frequency $\omega=\omega_{B,1}$, such that also $x_{{\rm opt},1}(\omega_{B,1})=0$. 
Specifically, for asymmetric systems, we find that when $y_1(\omega)=x_1(\omega)=0$ (that is, when $\omega=\omega_{B,1}$), the maximum of $\eta(\omega_{B,1})=\abs{\alpha_1(\omega_{B,1})}^2$ is obtained for 
$\Gamma_2=\Gamma_{{\rm opt},2}$ and $x_2(\omega_{B,1})=x_{{\rm opt},2}$ with
\begin{eqnarray}\label{Gammaopt2}
\Gamma_{{\rm opt},2}&=&\pt{1+x_{{\rm opt},2}^2}\
\pg{1+\frac{N_{T,2}}{N_{T,1}}\ \pq{\Gamma_{{\rm opt},1}(\omega_{B,1}) -1 } }
\ ,\\
x_{{\rm opt},2}&=&
\frac{y_2(\omega_{B,1})}
{
1+\frac{N_{T,2}}{N_{T,1}}\ \pq{\Gamma_{{\rm opt},1}(\omega_{B,1}) -1 } 
}\ ,
\end{eqnarray}
where according to Eq.~\rp{Gammaopt1} 
$\Gamma_{{\rm opt},1}(\omega_{B,1})=
\sqrt{\Gamma_\lambda\ \frac{N_{T,1}}{N_{T,2}} +1}$ 
and the corresponding transmission parameter is
\begin{eqnarray}\label{maxeta_asymm}
\eta(\omega_{B,1})=1-2\frac{\sqrt{\Gamma_\lambda+1}-1}{\Gamma_\lambda}\ ,
\end{eqnarray}
which is equal to Eq.~\rp{eq:eta_opt_N_opt} of the main text and is also valid for asymmetric systems.

In summary, the optimal parameters that simultaneously maximize the non-classicality ratio~\rp{criterion_asymm_opt} and the transmissivity~\rp{maxeta_asymm} of the gravity-induced channel is obtained for 
\begin{eqnarray}
\Delta_1&=&\omega_{B,1}=\omega \ , 
\\
g_1^2&=&\frac{\kappa_1}{4}\sqrt{
4\ \lambda^2
\frac{
\gamma_1\ N_{T,1}
}{
\gamma_2\ N_{T,2}
}
+\gamma_1^2
}\ ,
\\
\Delta_2&=&\omega_{B,1}
+\frac{\kappa_2\ \pt{\omega_{B,2}-\omega_{B,1}}
}{
\frac{N_{T,2}}{N_{T,1}}\ 
\sqrt{
4\ \lambda^2\ \frac{\gamma_2\ N_{T,1}}{\gamma_1\ N_{T,2}}+\gamma_2^2
}
-\gamma_2\ \frac{N_{T,2}-N_{T,1}}{N_{T,1}} 
} \ ,
\\
g_2^2&=&\frac{\kappa_2}{4}
\pt{
\frac{N_{T,2}}{N_{T,1}}
\sqrt{
4\ \lambda^2\ \frac{\gamma_2\ N_{T,1}}{\gamma_1\ N_{T,2}}+\gamma_2^2
}
+\gamma_2\ \frac{N_{T,2}-N_{T,1}}{N_{T,1}}
}
\pq{
1+\frac{
4\ \pt{\omega_{B,2}-\omega_{B,1}}^2
}{
4\ \lambda^2\ \frac{\gamma_2\ N_{T,2}}{\gamma_1\ N_{T,1}}+\gamma_2^2\ 
\frac{2\ N_{T,2}-N_{T,1}}{N_{T,1}}
}
}
-\frac{\kappa_2\ \gamma_2}{2}\ \frac{N_{T,2}-N_{T,1}}{N_{T,1}}
\ .
\end{eqnarray}

\section{Analysis of experimental requirements} \label{app:analysis_of_technical_requirements}

\subsection{Gravitational coupling constant and gravitational critical frequency}
To analyze any potential experimental implementation of the scheme proposed in the main text, it is necessary to make an explicit estimate of the gravitational coupling constant $\lambda$. In particular, it is important to expose its dependence on the mechanical frequency $\omega_B$. For example, it is intuitively clear that $\lambda$ should be a decreasing function of $\omega_B$ since trapping a particle in a stiff potential makes it less sensitive to external forces.
To make an explicit estimate of $\lambda$, we consider a simple model in which the two mechanical resonators correspond to two homogeneous spheres of radius $R$ that are positioned at an equilibrium distance $d$.
Their gravitational interaction energy is
\begin{equation}
U(x_1, x_2) = - G m^2 \frac{1}{d + x_2 - x_1} \simeq - \frac{Gm^2}{d} \left [ 1 + \frac{x_1 - x_2}{d} +\frac{(x_1 - x_2)^2}{d^2} + \dots \right ] = 
\text{local terms}  + \frac{Gm^2}{d^3} 2 x_1 x_2 + \mathcal O\left(\frac{(x_1-x_2)^3}{d^3}\right),
\end{equation}
where $x_j$ represents the displacement of the $j$-th sphere from its equilibrium position.
Neglecting third-order terms in the Taylor expansion and ignoring local terms (corresponding to a re-normalization of the local harmonic potentials), we obtain the desired quadratic interaction energy
\begin{equation}
V(x_1, x_2) =  \frac{Gm^2}{d^3} 2 x_1 x_2 = \frac{Gm\hbar}{d^3 \omega_B}  (b_1 + b_1^\dag)(b_2 + b_2^\dag), \label{eq:newtonian_approx}
\end{equation}
where we expressed the position variables $x_j$ in terms of the bosonic creation and annihilation operators $x_j=(b_j + b_j^\dag) \sqrt{\hbar/ (2 m \omega)}$.
Comparing the above equation with \eqref{eq:interaction}, we get an estimate for the gravitational coupling rate
\begin{equation}\label{eq:coupling_rate_lambda}
\lambda =  \frac{Gm }{d^3 \omega_B} \simeq \frac{G m}{8 R^3 \omega_B}= \frac{\pi}{6}  \frac{G m}{ V \omega_B}=\frac{\pi}{6}  \frac{G \rho}{\omega_B} \simeq \frac{w_G ^2}{\omega_B},
\end{equation}
where we assumed the minimum distance between the two spheres $d\simeq 2 R$ and introduced the {\it gravitational critical frequency} $w_G := \sqrt{\frac{\pi}{6} G \rho}$,
that, by construction, only depends on the geometry and mass density $\rho$ of the device and is independent of the trapping potential. For example, for two nearby gold spheres, we have a gravitational critical frequency of
\begin{equation}\label{eq:grav_w}
w_G  = \sqrt{\frac{\pi}{6} G \rho_{\rm Au}} \simeq 8.2 \times 10^{-4} {\rm Hz}.
\end{equation}
The physical interpretation of the above frequency is the following. For two harmonic oscillators of frequency $\omega_B \gg w_G $ (true for most real-world systems), the gravitational interaction can be considered as a weak perturbation. On the contrary, for  $\omega_B \approx w_G $, gravity is of the order of the harmonic force such that the approximation of two harmonic modes is broken.

\subsection{Optimal transmissivity, occupation number, and environmental critical frequency.}

It is common to express the quality of mechanical resonators in terms of the dimensionless quality factor $Q=\omega_B/\gamma$. Replacing $\gamma \rightarrow \omega_B/Q$ and $\lambda \rightarrow w_G^2 / \omega_B$ in \eqref{eq:eta_opt_N_opt}, we get:

\begin{equation}\label{eq:eta_opt_appendix}
\eta_{\rm opt}=\frac{2  Q^2 (w_G /\omega_B)^4}
{1+ \sqrt{1 + 4 Q^2 (w_G  /\omega_B)^4}+ 2  Q^2 (w_G /\omega_B)^4 }, \quad N_{\rm opt}= N_T=\left [\exp(\frac{\hbar \omega_{\rm B}}{k_B T}) -1\right]^{-1}=\left [\exp(\frac{ \omega_{\rm B}}{w_T}) -1\right]^{-1},
\end{equation}
where we introduced the {\it environmental critical frequency} $w_T := k_B T / \hbar$, that only depends on the environmental temperature $T$. The physical interpretation of $w_T$ is given by the Bose-Einstein statistics: For $\omega_B \ll w_T$, environmental noise obeys the classical law $N \simeq w_T/ \omega_B$; for $\omega_B \gg w_T$, environmental noise is exponentially suppressed by phonon quantization $N\simeq \exp(- \omega_B/w_T)$.
For example, at $T=1 {\rm mK}$, the environmental critical frequency is
\begin{equation}\label{eq:env_w}
w_T = k_B (1 \rm{mK}) / \hbar \simeq  1.3 \times 10^8 {\rm Hz}\ .
\end{equation}
%
The gravitational and environmental critical frequencies $w_G$ and $w_T$, introduced in Eq. \eqref{eq:grav_w} and \eqref{eq:env_w} respectively, provide a compact way of parametrizing the practical limitations of most experimental devices.  We derived their values assuming the idealized and simple model of two gold spheres at $T=1 {\rm mK}$. Ideally, we would like to have $w_G$ as large as possible (strong gravitational coupling) and $\omega_T$ as small as possible (low noise). For different experimental devices, $w_G$ and $w_T$ can change but their order of magnitude is likely to be similar to \eqref{eq:grav_w} and \eqref{eq:env_w} or worse. For example, by increasing the mass density as much as possible (e.g.~using osmium instead of gold) and by using different geometrical shapes, one may only increase $w_G$ by a small factor ($\lesssim 3$). In the same way, it is hard to imagine an optomechanical experiment at a temperature below the already optimistic hypothesis of $1 {\rm mK}$ and, therefore, it is hard to reduce $w_G$ below the value reported in \eqref{eq:env_w}.

Given that we have little hope for significant improvements in $w_G$ and $w_T$, we can fix them at the values estimated in Eq. \eqref{eq:grav_w} and \eqref{eq:env_w}, respectively. In this way, we are left with just two non-trivial experimental parameters: the quality factor $Q$ and the mechanical frequency $\omega_B$.
In the next subsection, we will explore potential experimental implementations over such a 2-dimensional parameter space.

\subsection{Non-classicality analysis in the $(\omega_B, Q)$ parameter space}

We can finally explore the 2-dimensional parameter space $(\omega_B, Q)$ to identify what are the potentially good regions for implementing the protocol proposed in this work, i.e., the regions in which gravity can induce a non-classical quantum channel according to the criteria $A$, $B$ and $C$, defined in the main text in Eqs.~(\ref{eq:criterion_A}-\ref{eq:criterion_C}).
According to the simple criterion derived in \eqref{eq:optimal_parameters} and reported in the main text in \eqref{eq:optimal_ratio}, the non-classical region is identified by the points $(\omega_B, Q)$ where
\begin{equation}\label{eq:parameters_criterion}
\frac{\lambda^2}{\gamma^2 N_T(\omega_B) (N_T(\omega_B) + 1)} =Q^2 \frac{w_G^4}{\omega_B^4}\frac{1}{N_T(\omega_B) (N_T(\omega_B) + 1)}= 4 Q^2 \frac{w_G^4}{\omega_B^4} [\sinh \left(\frac{\omega_B}{2 w_T}\right)]^2 > 1 \Leftrightarrow  2 Q \frac{w_G^2}{\omega_B^2} \sinh \left(\frac{\omega_B}{2 w_T}\right) > 1 \ .
\end{equation}
All the points in which the above condition is not satisfied correspond to the classical region in which the associated gravity-induced quantum channel is entanglement breaking and therefore unsuitable for the falsification protocol proposed in this work.

\begin{figure}[!h] 
    \includegraphics[width=0.9\linewidth]{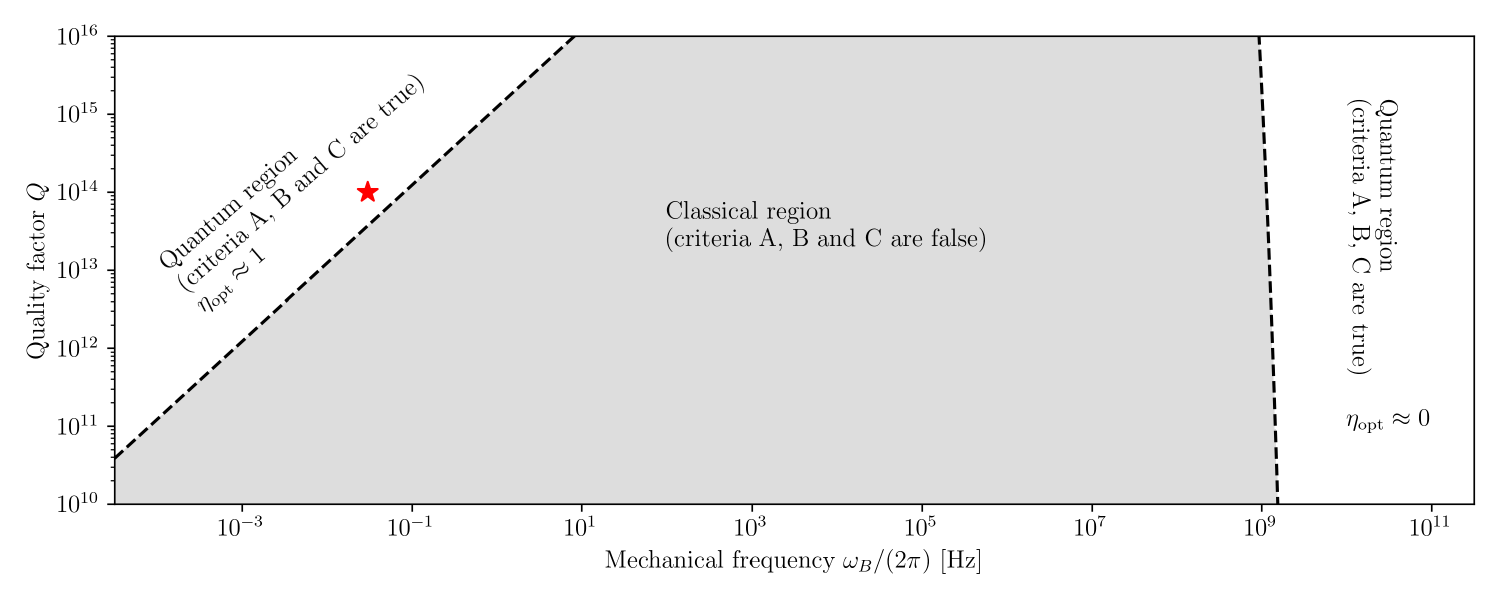}
    \caption{Non-classicality analysis of the gravity-induced channel in the $(\omega_B, Q)$ parameter space according to Eq.~\eqref{eq:parameters_criterion} and to the non-classicality criteria $A$, $B$ and $C$, defined in the main text in Eqs.~(\ref{eq:criterion_A}-\ref{eq:criterion_C}). This figure only assumes a linearized Newtonian force and quantum theory, without further experimental limitations apart from the gravitational and environmental critical frequencies reported in Eqs.~\eqref{eq:grav_w} and \eqref{eq:env_w}.
    In principle, there are two non-classical regions: one at very low frequencies (strong gravitational coupling) and one at very high frequencies (negligible thermal noise). However, as shown in Fig.~\ref{fig:transmissivity}, the transmissivity $\eta$ of the effective channel is $\approx 1$  in the low-frequency region, but it is extremely small ($\lesssim 10^{-23}$) in the high-frequency region. This fact makes the high-frequency region theoretically valid but experimentally problematic. The red star corresponds to the parameters of Fig.~\ref{fig:git}. }
    \label{fig:quality_frequency_space}
\end{figure}

The results are reported in Fig.~\ref{fig:quality_frequency_space}, where two quantum regions are identified at very low or very high frequencies. In the low frequency regime $\omega_B \ll w_T$, Eq. \eqref{eq:parameters_criterion} can be approximated to the simple condition
\begin{equation}\label{eq:low_freq_criterion}
  Q \frac{w_G^2}{w_T \omega_B} = \frac{w_G^2}{\gamma w_T}  \gtrsim 1,
\end{equation}
corresponding to a linear boundary in the logarithmic plot of Fig.~\ref{fig:quality_frequency_space}. Interestingly, Eq. \eqref{eq:low_freq_criterion} also implies that, at fixed critical frequencies $w_G$ and $w_T$, the only mechanical parameter that determines the non-classicality condition is the mechanical decay rate $\gamma$.
In the opposite limit of large frequencies $\omega_B \approx w_T$, Eq. \eqref{eq:parameters_criterion} is dominated by the exponentially exploding hyperbolic sine function, producing the approximately vertical boundary in the right part of Fig.~\ref{fig:quality_frequency_space}.

\begin{figure}[!h] 
\includegraphics[width=\linewidth]{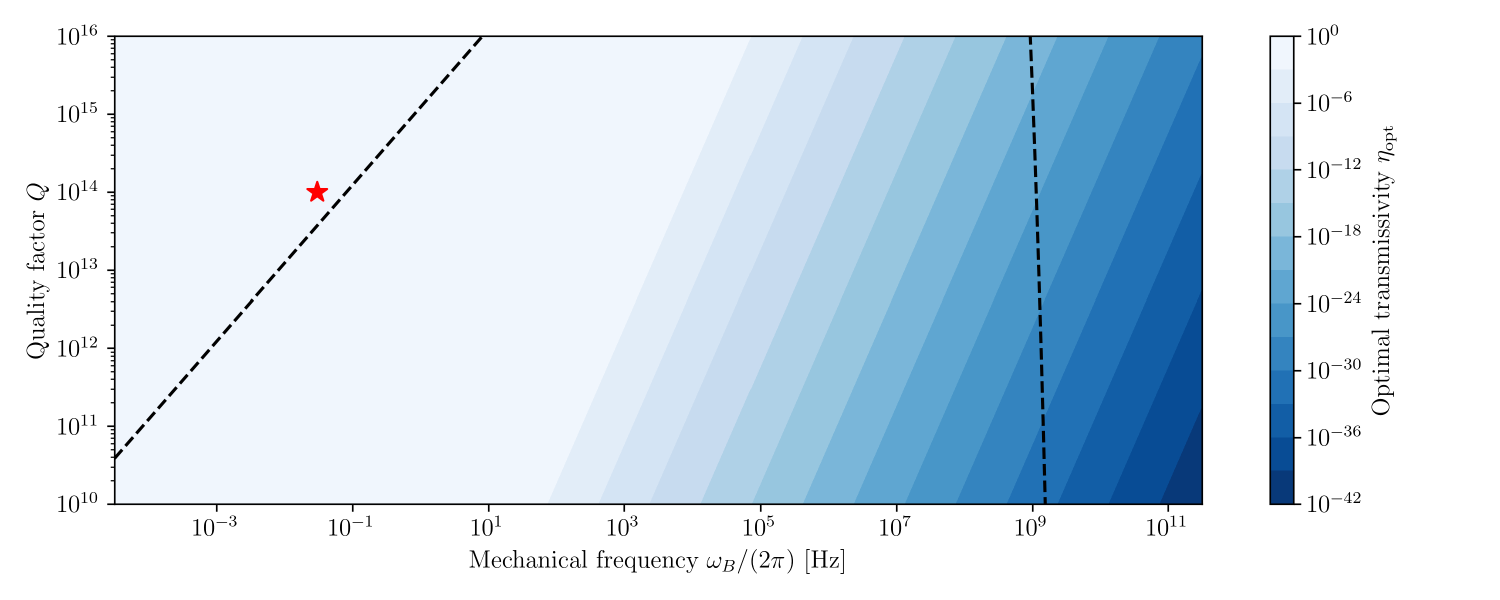}
    \caption{Effective optimal transmissivity $\eta_{\rm opt}$ of the gravity-induced optical channel in the $(\omega_B, Q)$ parameter space. 
    This figure assumes a linearized Newtonian force and an experimental apparatus characterized by the gravitational critical frequency reported in Eq. \eqref{eq:grav_w}.
    The transmissivity is $\approx 1$  in the low-frequency region, but it becomes extremely small in the high-frequency region, limiting experimental accessibility. The red star corresponds to the parameters of Fig.~\ref{fig:git}. The black dotted lines are the borders between the quantum and classical regions analyzed in Fig. \ref{fig:quality_frequency_space}.}
\label{fig:transmissivity}
\end{figure}

In the same parameter space used for  Fig.~\ref{fig:quality_frequency_space}, we can also plot the corresponding optimal transmissivity $\eta_{\rm opt}$ of the GIT channel, i.e., Eq.~\eqref{eq:eta_opt_appendix}. The result is shown in Fig.~\ref{fig:transmissivity}.

\newpage
\subsection{Minimum time duration of the experiment to resolve the GIT peak}

\begin{figure}[!h] 
\includegraphics[width=\linewidth]{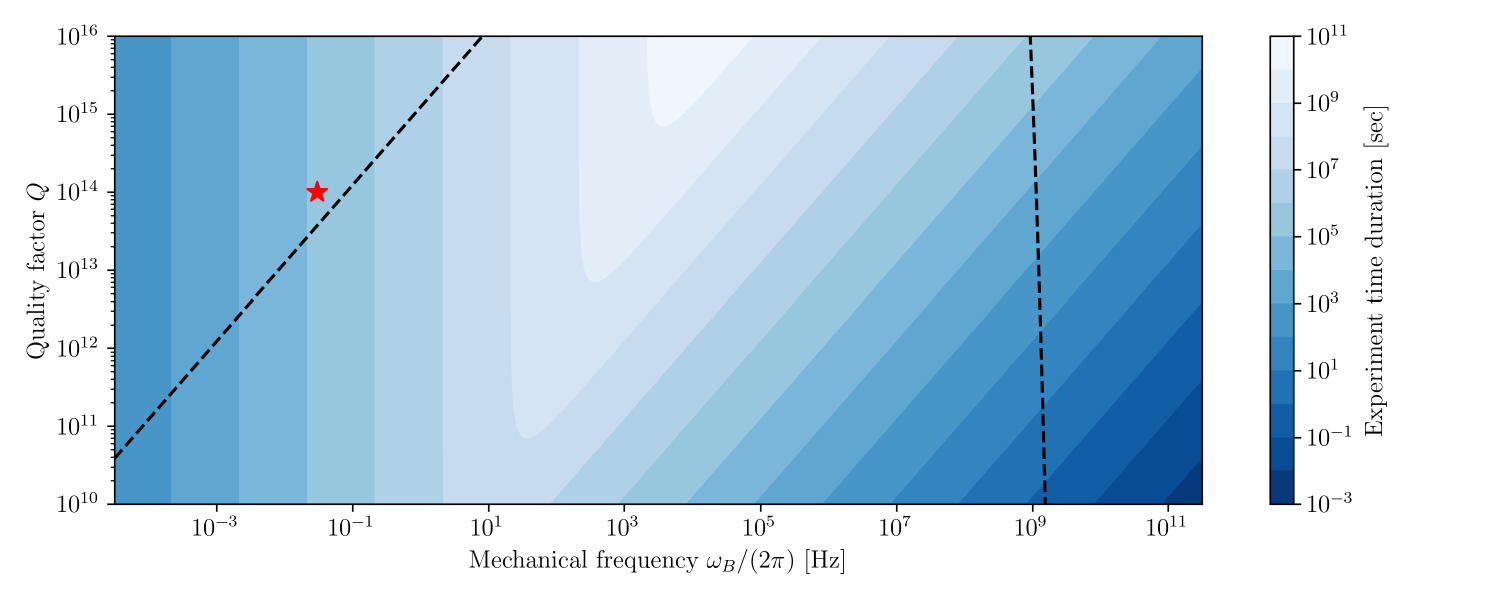}
    \caption{Minimum time required spectrally resolve the transparency window as estimated in Eq. \eqref{eq:experiment_duration}. This figure assumes a linearized Newtonian force and an experimental apparatus characterized by the gravitational critical frequency reported in Eq. \eqref{eq:grav_w}. The red star corresponds to the parameters of Fig.~\ref{fig:git}. The black dotted lines are the borders between the quantum and classical regions analyzed in Fig. \ref{fig:quality_frequency_space}.}
\label{fig:time_duration}
\end{figure}

The gravity-induced transparency discussed in the main text is only achievable in a narrow frequency window of the order of $\gamma^{\rm (eff)}_{\rm opt}$ estimated in \eqref{eq:linewidth}. As a consequence, one must consider a further  practical  experimental requirement: the overall duration $\tau$ of any experiment must be long enough to resolve the transparency bandwidth, i.e., we must have
\begin{equation}\label{eq:experiment_duration}
\tau \gtrsim  \tau_{\rm min} \approx \frac{1}{ \gamma_{\rm opt}^{\rm (eff)} } =\frac{1}{\gamma + \sqrt{\gamma^2 + 4 \lambda^2}}=\frac{Q/\omega_B}{1 + \sqrt{1 + 4 Q^2 w_G^4/\omega_B^4}} \ .
\end{equation}
This is a non-trivial technical requirement that must be considered when designing a real-world experimental implementation. Since gravity is a very weak interaction, the minimum time duration of the experiment in the non-classical regions can be of the order of hours or even days, as shown in Fig.~\ref{fig:time_duration}.

\subsection{Minimum effective transmissivity and minimum probe  power}

A further technical requirement is that the gravity-induced transmissivity cannot be too small since at least some photons of the input signal must be transmitted through the channel to make any meaningful experiment. Using equation \eqref{eq:eta_opt_appendix}, we can estimate and plot the optimal transmissivity in the full $(\omega_B, Q)$ parameter space as shown in Fig.~\ref{fig:transmissivity}.

\begin{figure}[!h] 
\includegraphics[width=\linewidth]{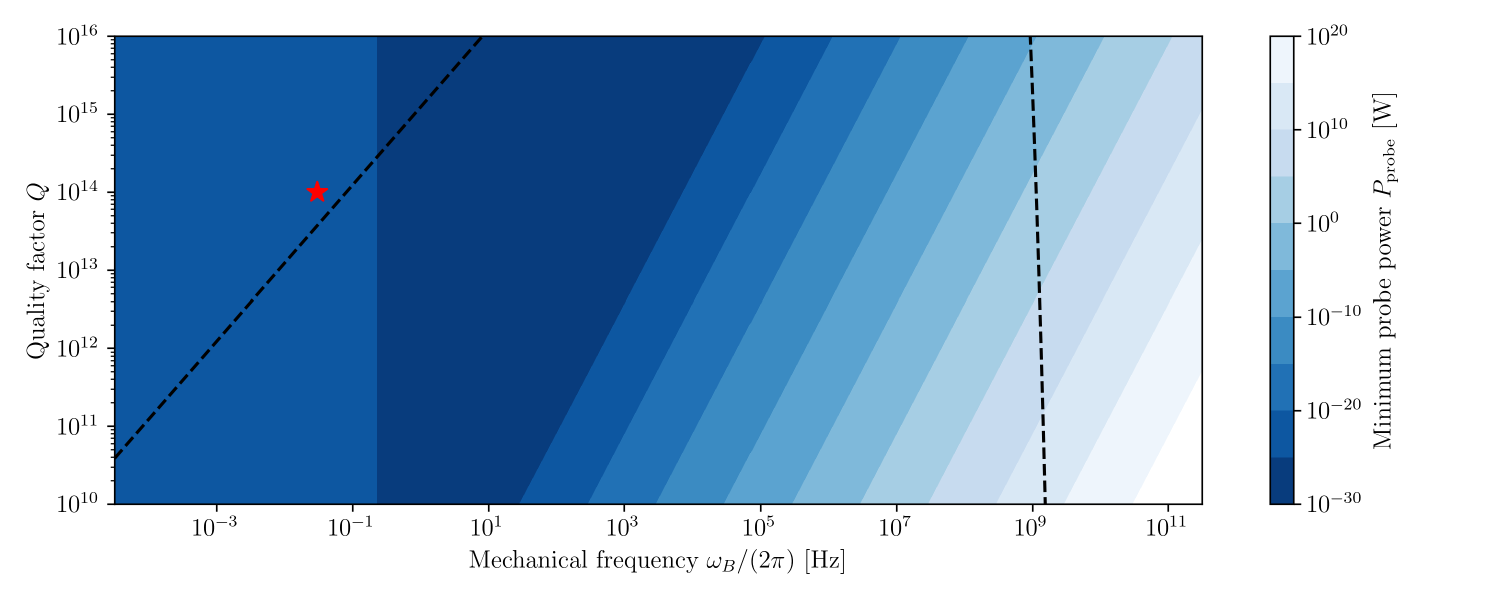}
    \caption{Minimum power of the probe signal such that at least one photon can be transmitted through the GIT channel in a time $\tau_{\rm min}$ given by \eqref{eq:experiment_duration}. The plot shows the lower bound estimated in Eq.~\eqref{eq:bound_on_power} for a cavity frequency of $\omega_A \approx 10^{15}$.
    This figure assumes a linearized Newtonian force and an experimental apparatus characterized by the gravitational critical frequency reported in Eq.~\eqref{eq:grav_w}. The red star corresponds to the parameters of Fig.~\ref{fig:git}. The black dotted lines are the borders between the quantum and classical regions analyzed in Fig. \ref{fig:quality_frequency_space}.}
\label{fig:min_power}
\end{figure}

Taking into account that the power of the input probe field is necessarily bounded (e.g.~it must be smaller than the laser pump power), we get the following lower bound:
\begin{equation}\label{eq:bound_on_eta}
\text{num. output photons} = (\text{num. input photons}) \eta_{\rm opt} \gtrsim 1  \Rightarrow 
\eta_{\rm opt} \ge  \frac{1}{\text{num. input photons}} \simeq  \frac{\hbar \omega_A}{P_{\rm probe} \tau},
\end{equation}
where $\tau$ is the time duration of the experiment.
Notice that this is not a limitation in the low-frequency regime where $\eta_{\rm opt} \approx 1$. However, this is a severe limitation for any experiment in the high-frequency regime.
If we fix the experiment time duration to $\tau_{\rm min}$ as defined in \eqref{eq:experiment_duration}, we can rewrite inequality \eqref{eq:bound_on_eta} as lower bound for the probe power that must be injected into the GIT channel in order to transmit at least a photon:
\begin{equation}\label{eq:bound_on_power}
P_{\rm probe} \ge  \frac{\hbar \omega_A}{\eta_{\rm opt} \tau_{\rm min}},
\end{equation}
The lower bound on the probe power is reported in Fig.~\ref{fig:min_power}. We observe that the lower bound is irrelevant in the low-frequency region but becomes very important in the high-frequency region.

\subsection{Comparison with the experimental requirements of gravity-induced entanglement experiments}\label{App_GITvsGIE}

The strongest advantage introduced by our proposal, compared to those requiring entanglement generation \cite{mari2016experiments, marletto2017gravitationally, bose2017spin, belenchia2018quantum, danielson2022gravitationally, marshman2020locality, krisnanda2020observable, matsumura2020gravity, christodoulou2023locally, galley2022no,poddubny2024nonequilibrium,lami2024testing,toccacelo2025benchmarks, kryhin2025distinguishable, miao2020quantum,  datta2021signatures, angeli2025probing, carlesso2019testing, hanif2024testing}, lies in the simplicity of the experimental test. Our protocol only requires assessing the non-classicality of a gravity-induced optical channel and, in practice, this can be done by sending single-mode coherent signals through the channel and  measuring the output signal. This is a simpler experiment than the generation and certification of gravity-induced entanglement.
A second advantage is that our protocol only requires controlling the system in its steady-state regime, a regime that is typically much easier to experimentally implement compared to a pulsed, time-dependent control.
A third advantage compared to many entanglement generation schemes is that such proposals are often focused on the generation of mechanical (or spin) entanglement, leaving its detection and certification as a further technical difficulty. In our setting,  the overall gravity-induced channel takes an optical input and produces an optical output, meaning that it can be characterized with standard quantum optical measurements (e.g., heterodyne detection).

Concerning experimental parameters, the requirements of our GIT protocol are similar to the requirements of entanglement-generation experiments. 
Gravity-induced entanglement is possible when its generation rate is larger than the decoherence rate of the system. Quantitative estimates of both quantities are given in \cite{bose2025massive} by Eq. (102) and Eq. (108), respectively. Imposing that their ratio must be larger than one, we get (assuming mechanical Brownian noise with spectral density  $S_{FF}=2 \gamma k_B T m$):
\begin{equation}
\frac{\text{Ent.\,  rate}}{\text{Dec.\, rate}}= 
\frac{2 G m^2 (\Delta x)^2 / (\hbar d^3)}{S_{FF}(\Delta x)^2/\hbar^2 } =
\frac{2 \hbar G m^2 }{S_{FF} d^3}  = \frac{2 \hbar G m^2 }{2 \gamma k_B T m d^3}=
\frac{\hbar G m }{\gamma k_B T d^3} \ge 1 . 
\end{equation}
Now let us compare the above inequality with the main inequality of our paper, according to which,  a non-classical gravity-induced channel is possible when:
\begin{equation}
\frac{\lambda^2}{\gamma^2 N_T (N_T + 1)} \simeq 
\frac{\lambda }{\gamma  N_T} = 
\frac{ G m / (d^3 \omega_B)}{\gamma k_B T / (\hbar \omega_B)}=
\frac{ \hbar G m }{\gamma k_B T d^3}
\ge 1 .
\end{equation}
Here, we used the explicit expression for the gravitational coupling rate  $\lambda$ that we derived in Eq.~\eqref{eq:coupling_rate_lambda}. We also used that, for low mechanical frequencies, the thermal occupation number is approximately $N_T \simeq k_B T / (\hbar \omega_B)$.
The final inequalities obtained for entanglement-generation experiments and for our gravity-induced channel experiment are the same. This means that there are no fundamental differences in terms of experimental parameters between the two types of experiments. On the other hand, as discussed above, there are important differences with respect to the conceptual and practical complexity of the experimental protocols.

\section{Linearized quantum gravity and the role of non-commutative field operators}\label{app_commutators}

The general protocol proposed in this work is model-independent with respect to the description of gravity. Indeed, by experimentally testing the non-classicality of the optical GIT channel, one can deduce the quantum nature of gravitational interaction without the necessity of assuming a specific model for it. 

However, it is instructive to describe our system within the specific framework of linearized quantum gravity \cite{bose2022mechanism, christodoulou2023locally}. This allows us to explicitly investigate the relationship between the non-classicality of the optical GIT channel and the non-commutativity of the gravitational field operators, thereby providing both a clarification and a self-consistency check of the results presented in the main text. In particular, we aim to show that—within linearized quantum gravity—the non-commutative nature of the field operators is essential for the transmission of quantum information from $S_1$ to $S_2$. In other words, if the gravitational field operators are replaced by classical \emph{c-numbers} (i.e., if the commutators are set to zero), it becomes impossible to transmit quantum information from $S_1$ to $S_2$.

Let us decompose the metric tensor as $g_{\mu \nu}({\bf r})= \eta_{\mu \nu} +h_{\mu \nu}({\bf r})$, where $\eta={\rm diag}(-1, 1, 1, 1)$ denotes the constant Minkowski background and $|h_{\mu \nu}( {\bf r})| \ll 1$ is a small perturbation. In this notation, the space-time indices $\mu, \nu$ can take values $\{1, 2, 3, 4\}$ and the signature is $(-, +, +, +)$.
We now promote the metric perturbation to a quantum operator \cite{bose2022mechanism}:
\begin{equation}
\hat{h}_{\mu \nu}({\bf r}) = \mathcal A \int d{\bf k} \sqrt{\frac{\hbar}{2 \omega_k (2 \pi)^3 }} [ {\hat P}_{\mu\nu} ({\bf k})^\dagger   - \eta_{\mu \nu} \hat P({\bf k})^\dagger]e^{-i {\bf k} \cdot {\bf r}} + \text{H.c.},   \label{eq:h_munu}
\end{equation}
where $\mathcal A= \sqrt{16 \pi G/c^2}$ and ${\hat P}_{\mu\nu}^\dagger, {\hat P}_{\mu\nu}, \hat P^\dagger,\hat P$ are the graviton creation and annihilation operators satisfying the bosonic commutation rules:

\begin{align}
[\hat{P}_{\mu, \nu}({\bf k}) , \hat{P}_{\lambda, \rho}({\bf k}')^\dagger] = (\eta_{\mu \lambda} \eta_{\nu \rho} + \eta_{\mu \rho} \eta_{\nu \lambda}) \delta({\bf k} - {\bf k}'), \nonumber \\
[\hat{P}({\bf k}) , \hat{P}({\bf k}')^\dagger] = - \delta({\bf k} - {\bf k}'). \label{eq:gravity_commutators}
\end{align}
Physically, the operators $\hat{P}_{\mu, \nu}({\bf k})$ represent spin-2 excitations, while $\hat{P}({\bf k})$ represent spin-0 excitations. If we are interested in the local field acting on a test mass, the overall field $\hat{h}_{\mu \nu}({\bf r})$ behaves as a simple linear combination of bosonic quantum modes, such that $\hat{h}_{\mu \nu}({\bf r})$ is the analogue of the electro-magnetic field locally acting on a test charge.
In this perturbative regime, the matter-field interaction, is governed by the following Hamiltonian \cite{bose2022mechanism, christodoulou2023locally}:
\begin{equation}
\hat H_{\rm int} = - \frac{1}{2} \int d{\mathbf r}  \hat{h}_{\mu \nu}({\bf r}) \hat{T}_{\mu \nu}({\bf r}). \label{eq:H-hT}
\end{equation}
where $\hat{T}_{\mu \nu}({\bf r})$ is the stress-energy tensor which, in our setting, describes the mass distribution of the mechanical degrees of freedom.
Before applying the above interaction Hamiltonian to the specific optomechanical systems considered in this work, we approximate the interaction to the case of a localized point-like particle.

For a single non-relativistic, point-like particle of mass $m$ and position operator $ \hat {\bf r}$, the dominant term of the stress-energy tensor is $\hat T_{00}({\bf r})= mc^2 \delta({\bf r} - {\hat {\bf r}})$, such that:
\begin{equation}
\hat H_{\rm int} \simeq - mc^2 \frac{\mathcal A}{2} \int d{\bf k} \sqrt{\frac{\hbar}{2 \omega_k (2 \pi)^3 }} [ {\hat P}_{00}({\bf k})^\dagger  + \hat P({\bf k})^\dagger]e^{-i {\bf k} \cdot {\hat{\bf r}}}  + \text{H.c.},
\end{equation}
Now, we assume the particle can only move by a small $\hat{\bf x}$ around an equilibrium position ${\bf r}_0$ (e.g. a trapped particle or a mechanical resonator). In this case, $\hat{\bf r} \simeq {\bf r}_0 + \hat{\bf x}$ and, neglecting high-frequency gravitons, we can linearize the exponential factor  $e^{-i {\bf k} \cdot {\hat{\bf r}}} \approx e^{-i {\bf k} \cdot {\bf r}_0} ( 1 - i {\bf k} \cdot {\hat{\bf x}})$ obtaining:
\begin{equation}
\hat H_{\rm int} \simeq  - mc^2 \frac{\mathcal A}{2} \int d{\bf k} \sqrt{\frac{\hbar}{2 \omega_k (2 \pi)^3 }} [ {\hat P}_{00} ({\bf k})^\dagger  + \hat P({\bf k})^\dagger]e^{-i {\bf k} \cdot {{\bf r}_0}} ( 1 - i {\bf k} \cdot {\hat{\bf x}})  + \text{H.c.}.
\end{equation}
Ignoring the irrelevant $\hat{\bf x}$-independent term, the interaction between the particle and the field reduces to:
\begin{align}
\hat H_{\rm int} \simeq  - \hat {\bf F} \cdot \hat{\bf x}, \label{eq:quantum_force}
\end{align}
where the Hermitian operator $\hat {\bf F}$ represents the quantum analogue of a classical gravitational force acting on the test particle and is given by:
\begin{align} \label{eq:quantum-force}
\hat {\bf F}=(\hat{F}_x, \hat{F}_y, \hat{F}_z)=mc^2\frac{\mathcal A}{2} \int d{\bf k} \sqrt{\frac{\hbar}{2 \omega_k (2 \pi)^3 }} [ {\hat P}_{00}({\bf k})^\dagger   + \hat P({\bf k})^\dagger]e^{-i {\bf k} \cdot {{\bf r}_0}} (-i {\bf k} ) + \text{H.c.},
\end{align}

\subsection{Replacing the quantum field with a classical c-number always generates a trivial classical dynamics}
 
Now we apply the above model to the optomechanical setup considered in this work, and we analyze the implications for the quantum nature of the associated gravitationally-induced transparency (GIT) channel. We are interested in showing that the non-commutative nature of the gravitational field is crucial to transmit quantum information from the input optical field of $S_1$ to the output optical field of $S_2$.

As a preliminary step of our analysis, we trace out $S_1$ and we only focus on the second optomechanical system $S_2$ and its local gravitational field. As we are going to show, this simplified picture is sufficient to understand the strong limitations implied by a classical description of the field. From Eq. \eqref{Hj} of the main text, we can write the corresponding Hamiltonian as:
\begin{align}
H_{2}(t)= \hbar \omega_B b_2^\dagger b_2 + \Delta a_2^\dagger a_2 + \hbar g (a_2 + a_2^\dagger) (b_2 + b_2^\dagger)- \hat{F}_x(t) \sqrt{\frac{\hbar}{2 m \omega_B}} (b_2 + b_2^\dagger), \label{eq:H-gravity}
\end{align}
where $b_2$ is the annihilation operator of the mechanical mode, $a_2$ is the annihilation operator of the optical mode, $\omega_B$ is the mechanical frequency, $\Delta$ is the optical detuning, and $g$ is the optomechanical coupling. The last term corresponds to the local quantum gravitational field acting as a time-dependent quantum force on the mechanical resonator according to Eq. \eqref{eq:quantum_force}.

If we replace the quantum field $\hat F_x(t)$ with a classical c-number $F_x(t)$, i.e., if we artificially set the commutators in Eq. \eqref{eq:gravity_commutators} to zero, $F_x(t)$ behaves as a real parameter and the Hamiltonian \eqref{eq:H-gravity} becomes quadratic in the bosonic operators and locally acting on system $S_2$ only. This means that the dynamics of the optomechanical system is necessarily a trivial Gaussian dynamics that can only generate a tiny subset of all possible quantum states, namely, those represented by a Gaussian Wigner function (assuming a Gaussian initial state for $S_2$). 
Moreover, since the output field $a_{\rm out_2}(t)$ exiting from the optical cavity depends linearly on $a_2(t)$ (see Eq. \eqref{eq:input-output}), its quantum dynamics is also constrained to be Gaussian. This is a strong limitation and directly implies the impossibility of transmitting (from $S_1$ to $S_2$) an arbitrary non-classical state such as a single-photon state $|1\rangle$, a cat state $\propto (\-|\alpha\rangle + |\alpha\rangle)$, or any quantum state whose Wigner phase-space representation is non-positive/non-Gaussian.

Note that a classical field $F_x(t)$ can, in principle, depend on the state of the first system $S_1$, such that the transmission of classical information is not forbidden. Classical information could, for example, be transmitted by encoding it in the amplitude or phase of the Gaussian output field $a_{\rm out_2}(t)$, establishing in this way a classical communication channel. What is instead impossible is the transmission of quantum information that, by definition, should allow the coherent transmission of arbitrary quantum states.

We have shown that we cannot replace $\hat F_x(t)$ with a classical c-number, i.e., that the non-zero commutators in Eq. \eqref{eq:gravity_commutators} are {\it necessary} to establish a non-classical GIT channel. However, is it possible to explicitly understand how such non-zero commutators can instead enable the transmission of quantum information? In other words, can we also show that the non-zero commutators in Eq. \eqref{eq:gravity_commutators} are also {\it sufficient} for transmitting quantum information (for sufficiently low thermal noise)?

We will use two alternative derivations. First, we'll show that a non-commutative gravitational field is able to induce arbitrarily complex non-Gaussian dynamics on system $S_2$ and is, therefore, able to bypass the discussed limitations of a classical field.
Secondly, we'll make a more explicit calculation including both $S_1$ and $S_2$ into the linearized quantum gravity model, demonstrating that a non-classical GIT channel is achievable.

\subsection{A non-commutative quantum field $\hat F_x(t)$ can drive the system $S_2$ into non-classical states}

In interaction picture with respect to the free field Hamiltonian  and the optomechanical Hamiltonian  $H_{OM}= \hbar \omega_B b_2^\dagger b_2 + \Delta a_2^\dagger a_2 + \hbar g (a_2 + a_2^\dagger) (b_2 + b_2^\dagger)$, the interaction term \eqref{eq:H-gravity} becomes linear in the bosonic operators of $S_2$: 
\begin{align}
H_{2}^I(t)=  - \hat{F}^I_x(t)  \left(\beta_+ b_+ e^{-i \omega_+ t} + \beta_+^* b_+^\dagger  e^{i \omega_+ t} +  \beta_- b_- e^{-i \omega_- t} + \beta_-^* b_-^\dagger  e^{i \omega_- t} \right),
\label{eq:H-gravity-interaction-picture}
\end{align}
where $b_{\pm}, b_{\pm}^\dagger$ describe the normal modes of the opto-mechanical Hamiltonian $H_{OM}$, $\omega_\pm$ are the associated frequencies, and $\beta_\pm \in \mathbb C$ are the decomposition coefficients of the position operator expanded in the basis of normal-modes, i.e.,  $\hat x_2=\left(\beta_+ b_++ \beta_+^* b_+^\dagger   +  \beta_- b_- + \beta_-^* b_-^\dagger \right)$. 
The associated time-evolution operator $U(t_0, t)$, from the initial time $t=t_0$ to the final time $t$, can be written in terms of the Magnus expansion:
\begin{align}
U(t_0, t)= \exp \left( \Omega_1(t_0, t) + \Omega_2(t_0, t)+ + \Omega_3(t_0, t) + \dots \right),
\end{align}
where 
\begin{align}
\Omega_1(t_0, t)&= -\frac{i}{\hbar} \int_{t_0}^{t}ds H_{2}^I(s), \\
\Omega_2(t_0, t) &= -\frac{1}{2\hbar^2} \int_{t_0}^{t}ds, \int_{t_0}^{s}ds'  [H_{2}(s), H_{2}^I(s')] \\
\vdots \\
\Omega_n(t_0, t) &=\frac{1}{n} \frac{- i}{\hbar} \int_{t_0}^tds  [H_{2}(s), \Omega_{n-1}(t_0, s')].
\end{align}

Now we can consider three different cases:

\begin{itemize}
\item If the operator $\hat{F}^I_x(t)$ is replaced by a classical c-number $F_x^I(t)$, it is easy to check that $\Omega_2(t_0, t)$ is also a c-number (since $[a_2, a_2^\dagger]=[b_2,b_2^\dagger]=1$) and, as a consequence,  $\Omega_n(t_0, t)=0$ for any   $n\ge 3$. In practice, since $\Omega_2(t_0, t)$ can only produce an irrelevant global phase, we can truncate the expansion to the first term $U(t_0, t)\propto \exp[\Omega_1(t_0, t)]$ corresponding to a phase-space displacement operator. A displacement operator can only trivially shift the quantum state of $S_2$ in phase space and is clearly unable to generate a non-classical quantum dynamics.

\item If $\hat{F}^I_x(t)$ is a bosonic field that is linear in creation and annihilation operators---such as Eq. \eqref{eq:quantum-force} rotated in interaction picture---then the commutator $[\hat{F}^I_x(t), \hat{F}^I_x(t')]$ is also c-number. This means that we can again write the evolution operator as $U(t_0, t)\propto \exp[\Omega_1(t_0, t)]$. However, in this case, $U(t_0, t)$ is not a displacement operator acting on $S_2$, but a unitary acting on the joint quantum state of the gravitational field and of $S_2$. This setting is compatible with the coherent transmission of quantum information \cite{serafini2023quantum}. In particular, if the gravitational field starts in a non-classical state or interacts with another non-classical quantum system, the final state of $S_2$ can also become arbitrarily non-classical.

\item If $\hat{F}_x(t)$ is an arbitrary quantum operator (e.g. an operator that may emerge from a non-linearized treatment of quantum gravity), the commutator $[\hat{F}^I_x(t), \hat{F}^I_x(t')]$ is, in general,  an non-trivial operator. This implies that all the terms of the Magnus expansion can be nonzero. The resulting dynamics would be non-Gaussian and therefore capable of generating non-classical quantum states.

\end{itemize}

We conclude that the non-commutative nature of the gravitational field can enable a non-classical dynamics for the reduced state of $S_2$ and is therefore {\it consistent} with the possibility of transmitting quantum information. On the contrary, a commutative field is {\it incompatible} with the transmission of quantum information. What remains to be investigated is whether the non-commutative nature of the field is also {\it sufficient} to generate a non-classical GIT channel between $S_1$ and $S_2$. This is the scope of the next subsection.

\subsection{Two optomechanical systems $S_1$ and $S_2$ coupled by the quantum gravitational field}

In the above analysis, we only studied the interaction of the gravitational field with a {\it single} mechanical degree of freedom,  specifically, the mechanical mode $b_2$ of the optomechanical system $S_2$ introduced in the main text.
However, the full setup described in the main text involves two optomechanical systems ($S_1$ and $S_2$) that are tuned to create an effective optical GIT channel linking the optical input of $S_1$ to the optical output of $S_2$.
A complete description of the entire $S_1$-gravity-$S_2$ system would be overly complex for our purposes, especially when including the optical input–output interactions and the thermal environment. For this reason, we split our analysis into two steps:
\begin{enumerate}
\item Deriving the effective interaction of two massive particles by perturbatively tracing out the gravitational field.
\item Showing that such effective interaction can induce a non-classical GIT channel between the optomechanical systems $S_1$ and $S_2$ considered in the main text.
\end{enumerate}
The combination of both steps explicitly demonstrates that the (non-commutative) gravitational field predicted by the theory of linearized quantum gravity can act as a mediator of quantum information.

Our starting point is the same gravitational field $\hat h_{\mu\nu}({\bf r})$ introduced in \eqref{eq:h_munu}  and the same interaction Hamiltonian introduced in Eq. \eqref{eq:H-hT}. There are two main differences compared to the previous analysis:

\begin{itemize}
\item We must take into account the free evolution Hamiltonian of the field \cite{bose2022mechanism}
\begin{equation}
\hat H_{\rm grav} = \int d{\bf k} \hbar \omega_{k} [\frac{1}{2}P_{\mu \nu}({\bf k})^\dagger P_{\mu \nu}({\bf k})- P({\bf k})^\dagger P({\bf k}) ]. \label{eq:H_grav}
\end{equation}

\item The dominant component of the stress-energy tensor is not a single delta function but the sum of two delta functions, associated the particle position operators $\hat{\bf r}_1$ and $\hat{\bf r}_2$:
\begin{equation}
T_{00}({\bf r}) = m_1 c^2 \delta({\bf r} - \hat{\bf r}_1) + m_2 c^2 \delta({\bf r} - \hat{\bf r}_2).
\end{equation}
\end{itemize}
We now consider the interaction Hamiltonian \eqref{eq:H-hT} as a small perturbation of the field Hamiltonian \eqref{eq:H_grav}. A similar approach was used in \cite{bose2022mechanism} for obtaining the Newtonian limit. Here, we apply a similar procedure for directly obtaining, in the limit of small displacement, an effective $\hat x_1$-$\hat x_2$ interaction between the two particles. 
The associated energy shift can be evaluated using second-order perturbation theory \cite{bose2022mechanism}:
\begin{equation}
\Delta \hat H_{\rm gravity}=\int d{\bf k} \frac{  \langle 0| \hat H_{\rm int}|{\bf k}\rangle \langle{\bf k}| \hat H_{\rm int}| 0\rangle}{\hbar \omega_k}, \label{eq:perturbation}
\end{equation}
where $|{\bf k}\rangle = \left( \hat P_{\mu \nu}({\bf k})^\dagger + \hat P({\bf k})^\dagger \right) |0\rangle$ are  single-excitation states of the gravitational field with energy $\hbar \omega_k$ and $|0\rangle$ is the vacuum state of the field. 
From the commutation rules \eqref{eq:gravity_commutators} of the gravitational creation and annihilation operators, it can be shown (see e.g. footnote 8 of \cite{bose2022mechanism}) that:
\begin{equation}
\langle {\bf k} |  \hat P_{\mu \nu}({\bf k'})^\dagger + \hat P({\bf k'})^\dagger |0\rangle= \delta({\bf k} - {\bf k'}). \label{eq:delta_kkp}
\end{equation}
Inserting the interaction Hamiltonian \eqref{eq:H-hT} into \eqref{eq:perturbation} and using both  \eqref{eq:h_munu} and \eqref{eq:delta_kkp}, we get:
\begin{equation}
\Delta \hat H_{\rm gravity} = m_1 m_2 c^4 \frac{\mathcal A^2}{4} \int d{\bf k} \frac{1}{2 \omega_k^2 (2\pi)^3}(e^{i {\bf k} \cdot \hat {\bf r}_1} +  e^{i {\bf k} \cdot \hat {\bf r}_2}) (e^{-i {\bf k} \cdot \hat {\bf r}_1} +  e^{-i {\bf k} \cdot \hat {\bf r}_2}).
\end{equation}
Now, as we did for the single-particle case, we assume both particles can have small displacements $\hat {\bf x}_j$ around their respective equilibrium positions ${\bf r}_j$, such that $e^{\pm i {\bf k} \cdot \hat {\bf r}_j}\simeq e^{\pm i {\bf k} \cdot {\bf r}_j} (1 \pm i  {\bf k} \cdot \hat {\bf x}_j)$. Within this approximation and neglecting the irrelevant local terms (they can be absorbed in the single-particle Hamiltonians), we are left with the following interaction potential:
\begin{equation}
\Delta \hat H_{\rm gravity} = m_1 m_2 c^4 \frac{\mathcal A^2}{4}  \int d{\bf k} \frac{1}{2 \omega_k^2 (2\pi)^3}(e^{i {\bf k}\cdot ({\bf r_1} - {\bf r_2})} + e^{-i {\bf k}\cdot ({\bf r_1} - {\bf r_2})})  ( {\bf k} \cdot \hat{\bf x}_1)({\bf k} \cdot \hat{\bf x}_2) .
\end{equation}
Now, we assume that the particles are constrained to move along the same direction, say along the $x$ axis. In this case we can make the substitution ${\bf r}_j\rightarrow (r_j, 0, 0)$ and $\hat {\bf x}_j\rightarrow(\hat x_j, 0, 0)$

\begin{equation}
\Delta \hat H_{\rm gravity} = m_1 m_2 c^4 \frac{\mathcal A^2}{4}  \int d{\bf k} \frac{1}{2 \omega_k^2 (2\pi)^3}(e^{i k_x \cdot ({r_1} - {r_2})} + e^{-i {k_x}\cdot ({r_1} - { r_2})} ) k_x^2   \hat{x}_1 \hat{x}_2 .
\end{equation}
Using $\omega_k=|{\bf k}| c$, $\mathcal A= \sqrt{16 \pi G/c^2}$, and the Fourier integral $\int d {\bf k} e^{i k_x r} (k_x/|{\bf k}|)^2= -d^2/dr^2 (2 \pi^2/r)=(2 \pi)^2/r^3$, we get
\begin{equation}
\Delta \hat H_{\rm gravity} = m_1 m_2 c^2 \frac{\mathcal A^2}{4} \frac{1}{2 \pi} \frac{1}{|r_1-r_2|^3} \hat{x}_1 \hat{x}_2 =
 \frac{2 m_1 m_2 G}{|r_1-r_2|^3} \hat{x}_1 \hat{x}_2.
\end{equation}
This is the quadratic coupling with exactly the same prefactor that we explicitly considered in this work, specifically in Eqs. \eqref{eq:newtonian_approx} and \eqref{eq:interaction}.
This means that the second step of this section---i.e., demonstrating that the above effective interaction can be used to induce a non-classical quantum channel between 
$S_1$ and $S_2$---exactly corresponds to the same analysis already performed in the main text and in the Supplemental Material of this work. 

Therefore, we can finally conclude that---within the framework of linearized quantum gravity---the non-commutative nature of the gravitational field is exactly the key non-classical feature that enables the possibility of transmitting quantum information.
\end{document}